\documentclass[apj,twocolumn]{aastex631}
\usepackage{epsfig}
\usepackage{natbib}
\usepackage{amsmath}
\usepackage{hyperref}
\usepackage{mathrsfs}
\usepackage{graphicx}
\usepackage{xfrac}
\newcommand{\meanchi}{\textlangle$\chi$\textrangle}

\newcommand{\kms}{km~s$^{-1}$}
\newcommand{\Msun}{M_{\odot}}
\newcommand{\Lsun}{L_{\odot}}

\newcommand{\p}{^\prime}
\newcommand{\kmsMpc}{km~s$^{-1}$~Mpc$^{-1}$}

\def\dcolheadv#1{\multicolumn{1}{c|}{$\vrule depth 6pt height 12pt width 0pt\relax#1$} \ignorespaces}

\begin{document}
\title{Galaxy flows within 8,000~\kms\ from Numerical Action methods}

\author{Edward J.Shaya}
\affil{Astronomy Department, University of Maryland, College Park, MD 20742, USA: eshaya@umd.edu}
\author{R. Brent Tully}
\affil{Institute for Astronomy, University of Hawaii, 2680 Woodlawn Drive, Honolulu, HI 96822, USA}
\author{Daniel Pomar\`ede}
\affil{Institut de Recherche sur les Lois Fondamentales de l'Univers, CEA Universit\'e Paris-Saclay, 91191 Gif-sur-Yvette, France}
\author{Alan Peel}
\affil{Astronomy Department, University of Maryland, College Park, MD 20742, USA: eshaya@umd.edu}

\shorttitle{Galaxy flows within 8,000~\kms}
\shortauthors{Shaya et al.}
\accepted{to ApJ}

\begin{abstract}
The trajectories since $z=4$ of systems of galaxies (`halos') with $cz < 8,000$ km s$^{-1}$ are found through Numerical Action reconstructions.  A set of 9,719 halos from a 2MASS group catalog and {\it Cosmicflows-3} catalogs are given attention. Present distances are adjusted to minimize departures from observed redshifts. For those with the most precisely determined distances, compromises are made between distance and redshift agreement.  $H_0$ is varied from 69 to 77~ km s$^{-1}$ Mpc$^{-1}$ with $\Omega_m$ set by the baryon acoustic oscillation constraint from the Planck Satellite.   A best fitting amplitude of the mass-to-light relation is found. A uniform density associated with the interhalo medium accounts for the matter not in halos.  The solution paths provide the histories of the formation of the nearby large structures and depict how the voids emptied.   Assuming no local over/underdensity, the best model has $H_0=73$ \kmsMpc\ with nearly the same density arising from interhalo matter (IHM) as from halos.   We examine local over/underdensities by varying the IHM density and find a valley of best fit models along $H_0 = 73.0 (1 + 0.165\delta)$  \kmsMpc.  Friedmann models with distinct densities internal and external to the study region give a similar relationship. The fraction of matter in the IHM seen in n-body simulations roughly matches that in our $H_0=72$ scenario.  Videos have been created to visualize the complexities of formation of large-scale structures. Standard n-body calculations starting from the first time-steps as tests of the NAM solutions, and continue until cosmic scale factor $a=2$  provide glimpses into the future.

\end{abstract}

\section{Introduction}

A grand goal in cosmology would be to know the full trajectories of all the pieces of all the galaxies within a volume of the universe around us large enough to be considered representative.  Imaginary macrocosms can be simulated \citep{2006Natur.440.1137S, 2018MNRAS.475..676S, 2014MNRAS.444.1518V, 2015MNRAS.446..521S}.  There are simulations based on constrained initial conditions that can generate major clusters ($>10^{14} \Msun$) and voids at locations that approximate the observed nearby universe \citep{2003ApJ...596...19K, 2016MNRAS.455.2078S}.  However stochastic processes prevent the accurate placements of entities on scales of small groups and individual galaxies that accurately mimic the real world.

\citet{1989ApJ...344L..53P} introduced an alternative way forward.  The comoving coordinate orbits of particles follow paths that are extrema or saddle points of the action, $S$, the integral of the Lagrangian, $\mathscr{L}$, over time, $t$:
\begin{equation}
    \delta S = \delta \int_0^{t_0} \mathscr{L} dt = 0
\end{equation}
where for collisionless particles the Lagrangian is the comoving kinetic minus the comoving potential energy in the system:
\begin{equation}
    \mathscr{L} = T - V
\end{equation}
The boundary constraints include the requirement that all initial peculiar velocities are small.  There have been various studies employing numerical action methods to model galaxy orbits \citep{1995ApJ...454...15S, 2001ApJ...554..104P, 2013MNRAS.436.2096S, 2017ApJ...850..207S}.  
\cite{2000MNRAS.313..587N}, in particular, studied the accuracy of NAM and found that it reconstructs velocities faithfully in high density regions where deviations from the Zel'dovich solution are large.

The ability to model the non-linear dynamics of infall regimes is an important advantage of the numerical action method.   On scales of a few Mpc, most galaxies and groups are, at least slightly, in a non-linear interaction with nearby masses. The non-linear regime infall of test particles provide strong constraints on the masses of major attractors over a wide range of radii. The ubiquitous voids, being underdensities of $\delta = (\rho/\rho_m - 1) \sim -1$, obviously generate motions at their peripheries (i.e., most everywhere) not in the linear regime, defined as $|\delta| \ll 1$.   

The orbits are followed from an early time, here from $z=4$, age of the universe $\sim$1.6 Gyr.  Back then, the entities we see today were in many pieces.  The premise that is adopted is that the orbits trace the centers of mass of those many pieces that have assembled into the current halos, and the center of mass is unaffected by the details of the merger tree.  
Limitations of orbit reconstructions must be acknowledged.  For one thing, there is usually insufficient information to untangle the orbits through multiple shell crossings in collapsed regions.\footnote{The case of the Local Group might be the exception \citep{2013MNRAS.436.2096S}, especially as proper motions become more reliably established.}
Hence, the units that will be considered in the construction of our sample catalog are collapsed halos that often include a multitude of galaxies. The dominion of a halo extends to roughly the virial radius.  Entities that are infalling, not yet at shell crossings, are kept distinct.

The viability of the model depends on whether its sample catalog reasonably captures the distribution of matter.  Calculations become oppressive if the number of entities is large, but a trial solution of $\sim$10,000 entities can now be found by high end desktop computers in less than a day. There can be considerable work afterward, though, untangling tight interactions and choosing among multiple path solutions.

There should be as close to complete three-dimensional coverage as possible around the origin, in this case the Local Group.  However, there is missing information in the zone of obscuration on the Galactic equator.  Given the computational penalty for carrying excessive entries, very small galaxies and groups are excluded.   Then the matter outside of the luminous halos must be taken into account.  Large n-body simulations show that the amount of matter in sheets, filaments, halos too small to form and retain significant numbers of stars, or free streaming is comparable to the amount of matter in luminous bound systems.  This assertion is supported by a study of the unbound particles in the Millennium Run n-body calculation \citep{2015MNRAS.452.1779S}. Studies of voids show that they are not completely empty of mass. They are threaded by networks of matter with average density at a few tenths of mean density \citep{2003MNRAS.344..715G}.   

There are two ways that entities (virialized halos or individual galaxies) can enter the list considered for orbit reconstruction.  The first is because they cross an intrinsic luminosity threshold that infers they are important constituents of the mass distribution.  The second is because they have measured distances that are considered accurate above a threshold.  
The more entries with accurately known distances the better.  These are test particles responding to and informing us of the mass distribution.   

The numerical action methodology generates orbits that are physically possible within the constraints of the distribution of constituents.  However, there may be a number of plausible options, particularly in crowded regions.    
The specifics of individual orbits are tenuous.  However, mass estimates of the major components and the overall density are reasonably robust (within the specifics of the assumed cosmology).  Also, the overall flow patterns are robust.

The current study is an extension of \citet{2017ApJ...850..207S} where the volume considered was limited at 38~Mpc $\sim$2,850~\kms\ and 1,382 orbits were followed.  That volume encompassed the traditional Local Supercluster \citep{1953AJ.....58...30D} with the Virgo Cluster the dominant mass concentration and the Local Void a dominant dynamical factor \citep{2019ApJ...880...24T}.  In this new study, orbits are reconstructed to a limit of 8,000~\kms\ in the CMB frame, $\sim$107~Mpc, a volume 22 times greater, and 9,719 entities are given attention.  

\begin{figure}[ht]
   \centering
    \includegraphics[width=0.47\textwidth]{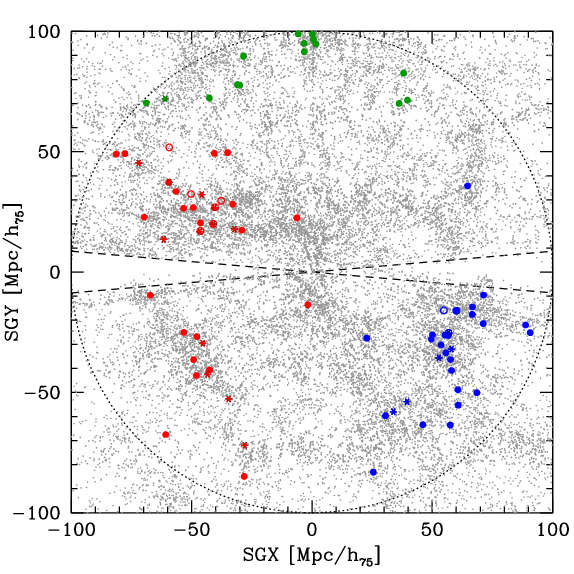}
    \caption{\textbf{Distribution of 2MASS galaxies within 100~Mpc} in a projection onto the supergalactic equator. Major clusters are given colors; red if in the Laniakea Supercluster; blue if associated with the Perseus-Pisces filament; and green if in the Great Wall. The zone of Galactic obscuration projects into the $\pm 5^{\circ}$ wedge.}
    \label{fig:biggp}
\end{figure}

There were two considerations in establishing the volume limit.  One is the practical reason that information on distances is thinning out around this limit in the base {\it Cosmicflows-3} catalog \citep{2016AJ....152...50T}.  The other is related to the known structural elements in our vicinity.  The 8,000~\kms\ limit  (CMB frame) comfortably encloses both the so-called Great Attractor region \citep{1987ApJ...313L..37D, 1988ApJ...326...19L}, the core of Laniakea Supercluster \citep{2014Natur.513...71T}, and the Perseus-Pisces complex \citep{1988lsmu.book...31H}.  See Figure~\ref{fig:biggp}.  These two high density regions are given particular attention in this study.  A third complex, the Great Wall \citep{1986ApJ...302L...1D}, crosses the 8,000~\kms\ boundary and any analysis of it is tentative.  Aside from the Great Wall structure, the 8,000~\kms\  boundary generally runs through uncomplicated void-like regions.  

In \S\ref{sec:data} there will be a presentation of the input data to the model.  In \S\ref{sec:nam} there will be a discussion of the numerical action methodology and the cosmology assumptions. \S\ref{sec:results} presents statistical results from the numerical action study.  Cosmographic representations of results are given in \S\ref{sec:flows}.  In \S\ref{sec:future} the evolution of structure is followed into the future. There are brief conclusions in \S\ref{sec:summary}. 

\pagebreak
\section{Data}
\label{sec:data}

The numerical action calculation minimally needs constituents that provide good representation of the large scale clustered mass distribution through apparent brightnesses and redshifts.  Most entities we will consider have redshifts with high S/N and sufficient accuracies.   In addition, to invoke goodness criteria, good distances are needed for a substantial fraction of the entities, and these should be well distributed over the entire sample volume. This study has awaited the availability of the copious assemblage of distance determinations given in the {\it Cosmicflows-3} compilation \citep{2016AJ....152...50T}.   

\subsection{Mass distribution}
\label{sec:masses}

The description of the mass distribution draws on two galaxy group catalogs. One draws on the collection of redshifts \citep{2012ApJS..199...26H} for 2MASS, the Two Micron All Sky Survey \citep{2006AJ....131.1163S} that is quasi-complete to $K_s=11.75$ outside a small strip along the Galactic plane at $|b|< 5^{\circ}$ . The creation of a 2MASS-based group catalog began with a study of selected well studied groups and clusters over the wide range of masses from $10^{11}$ to $10^{15} \Msun$ \citep{2015AJ....149...54T}.  That study provided scaling relationships between luminosity, group velocity dispersion, $\sigma_p$, the characteristic radius of second turnaround, $R_{2t}$, and group mass. 

The scaling relationships provide the tools to build a group catalog based on the 2MASS redshift compilation to $K_s=11.75$ \citep{2015AJ....149..171T}.  The 2MASS compilation is close to optimal because of its photometric integrity and its sensitivity to older stellar populations that represent most of the baryonic mass.  Most of the mass in stars in the domain of the local universe that concerns us (outside the gap on the Galactic plane) is probably well represented in the 43,000 galaxies in the 2MASS redshift catalog.

\begin{figure}[ht]
   \centering
    \epsscale{1.21}
    \plotone{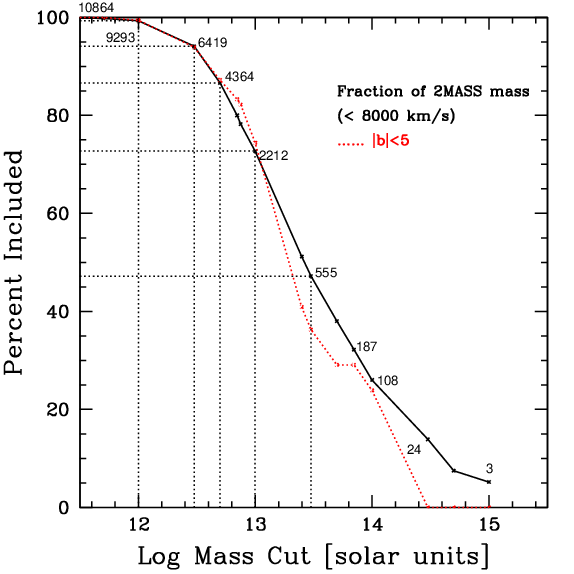}
    \caption{\textbf{Percent inclusion of mass associated with the 2MASS $K_s<11.75$ redshift catalog as a function of the mass cut on groups.}  Numbers at intervals along the black curve give the cumulative numbers of groups as a function of mass.  The red curve gives equivalent information for the normalized mass within the $\pm5^{\circ}$ obscured zone.}
    \label{fig:masscut}
\end{figure}

The entries in the group catalog are collapsed halos that can range from rich clusters with hundreds of associated 2MASS galaxies down to individual isolated galaxies.  
To begin the construction of the catalog, the intrinsically brightest galaxy in the 2MASS collection (assuming Hubble distances) was given attention and gathered within its domain all other galaxies as expected from the scaling relationships.  
Specifically, galaxies are linked if their velocities agree within $2\sigma_p$ and their projected positions lie within $R_{2t}$.  
The inclusion of new members dictates iterations based on the increased luminosity.  
The masses statistically associated with halo luminosities are taken to grow with a weak power of luminosity: $M \propto L^{1.15}$.  
Once the composition of the halo associated with the brightest galaxy is secured, the most luminous galaxy in the remainder of the 2MASS catalog is given consideration, and so on, until all the galaxies have been linked to a halo, including halos of a single member.

The cumulative mass fraction as a function of halo mass is shown in Figure~\ref{fig:masscut}.  It is seen that almost all the mass in collapsed halos is contained within halos above $10^{12}~\Msun$.  There are $\sim$10,000 such halos within 8,000~\kms. 

There are details, such as arise when halos are in close proximity and may be merged or split, and instances of what must be high velocity members.  These complexities are discussed by \citet{2015AJ....149..171T}.  
An issue of relevance is adjustments that are made for lost luminosity as a function of distance. 
 The assumption is made that the summed luminosity in radial shells after adjustment should be constant. A formula for a correction factor was empirically determined to achieve this criterion (see Fig.~5 and Eq.~9 in \citet{2015AJ....149..171T}).

By 10,000~\kms\ the apparent magnitude cutoff of $K_s=11.75$ for the catalog is in the vicinity of the absolute magnitude $M^{\star}_{K_s}$, at the exponential break in the Schechter function  \citep{1976ApJ...203..297S} (taking $M^{\star}_{K_s} =-23.55 +5{\rm log}$h, h = $H_0/100$ \kmsMpc , and slope parameter $\alpha_{K_s}=-1.0$).
Correction factor adjustments are below a factor two up until this velocity but then rapidly increase with redshift.  Corrections for missing light are moderate within our limit of 8,000~\kms.

The 2MASS based group catalog is less than ideal nearby because the space density of the high surface brightness galaxies that enter the 2MASS compilation is low and because redshifts break down as a way of establishing relative galaxy distances.  
Nearby, within 3,500~\kms, there is the preferable group catalog of \citet{2017ApJ...843...16K}.  The groups in their catalog are based on the same scaling relationships.  It includes all 15,000 galaxies that were known within the velocity limit.  
For many small galaxies, $K_s$ magnitudes had to be inferred from optical fluxes \citep{2003AJ....125..525J, 2005AstL...31..299K}.  
Measured distances were used when available, allowing the rejection of group interlopers and giving improved clarity to group memberships down to $10^{10} \Msun$.

There remains the problem of the zone of obscuration, a gap in the 2MASS compilation of 9\% of the sky.  Locally, within 2,850~\kms, this problem was addressed by \citet{2017ApJ...850..207S} with the addition of sources from the main 2MASS catalog \citep{2000AJ....119.2498J} at $|b|<5^{\circ}$ combined into groups.  Consideration of blind HI surveys \citep{1987ApJ...320L..99K, 2016AJ....151...52S} gave assurance that important obscured structures were given representation and these same elements are included in this new study.

Beyond 2,850~\kms\ out to 8,000~\kms, the $\pm5^{\circ}$ Galactic equatorial gap widens in physical space and there must be compensation for what is missed.  There are known clusters and individual galaxies at these low latitudes from x-ray, infrared, and radio surveys and follow up redshift observations \citep{2005RvMA...18...48K, 2007ApJ...662..224K}.  These entities are given reasonable mass estimates and included.  However, these known pieces are insufficient, as judged by the subsequently averaged density at low latitudes.  To compensate, their masses are amplified by a factor four to recover densities associated with halos roughly equal to values in unobscured regions.  It is supposed that their inflated masses account for missing mass in their proximity.  Clearly, the orbits found for them and those adjacent should be considered unreliable.

\subsection{Accurate distances}

Galaxy distance measures are drawn from the {\it Cosmicflows-3} compilation \citep{2016AJ....152...50T}.  Contributions come from six methodologies.  Two were foundational for the absolute scale: the Cepheid period-luminosity relation \citep{1912HarCi.173....1L, 2019ApJ...876...85R} and the tip of the red giant branch method \citep{1993ApJ...417..553L, 2007ApJ...661..815R}.  The latter, TRGB, is particularly noteworthy because the $\sim$500 distances with the $\sim$5\% accuracy it has provided results in an outstandingly precise view of structure and peculiar motions within 10 Mpc.

Two other methodologies provide distance estimates in a wide $30-300$~Mpc range that are accurate at the level of 10\% but are limited in number: surface brightness fluctuations \citep{1988AJ.....96..807T, 2015ApJ...808...91J} and Type Ia supernovae \citep{1993ApJ...413L.105P}.  It can be anticipated that contributions from these techniques will be increasingly relevant in the future.

It is the two remaining methodologies that contribute the most numerically although their accuracies per galaxy are only $20-25\%$: the early-type fundamental plane \citep{1987ApJ...313...42D, 1987ApJ...313...59D, 2012MNRAS.427..245M} and spiral galaxy luminosity-linewidth \citep{1977A&A....54..661T, 2020ApJ...896....3K} relations.  These techniques contribute distances to $\sim$97\% of the 17,647 entries in {\it Cosmicflows-3}.  Within 8,000~\kms, the concern of this study, the coverage around the sky outside the zone of obscuration is reasonably uniform.\footnote{At $8,000- 15,000$~\kms\ the coverage in {\it Cosmicflows-3} is greater in the celestial south because of the important contribution from the fundamental plane component of the Six Degree Field survey \citep{2014MNRAS.445.2677S}.} 

\begin{figure}[ht]
   \centering
    \includegraphics[width=\linewidth]{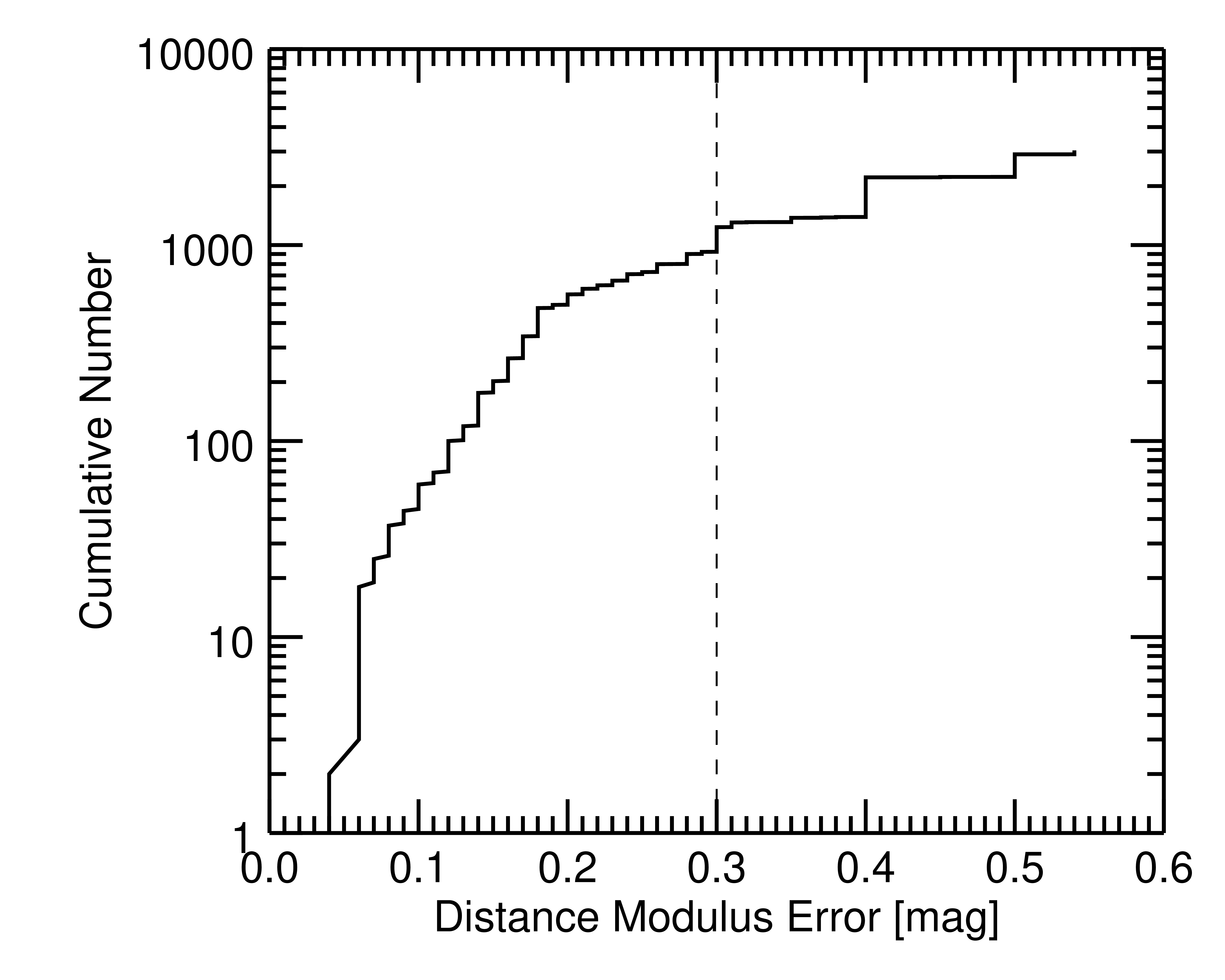}
    \caption{\textbf{Cumulative number of distance modulus errors for halos in our input catalog.}}
    \label{fig:edm}
\end{figure}

The accumulation of distance estimates within groups and clusters is important beyond the linkage of methodologies.  The separate distance moduli determined by different methods for a given galaxy or for members of a group can be averaged, with appropriate weights. 
The entities in our catalog are assigned velocities averaged over all known members and distances averaged in the modulus over all members with suitable measurements.
Consequently, the distances to some halos have high statistical accuracy.  
In the extreme, for the Virgo Cluster there are 170 measures.  
The statistical error on the cluster distance is 1\%.  Systematic errors dominate.  
Figure~\ref{fig:edm} provides the number of halos with better than a given uncertainty in distance modulus.  
The 924 halos with distance errors $\mu_e < 0.30$ mag (left of vertical dotted line) nominally satisfy our criterion for model goodness.

\subsection{Influences beyond 8,000 km/s}
\label{sec:external}

It can be expected that external tidal influences are a concern for the present study, particularly toward the edges of the sample and perhaps for the overall dipole motion within the volume.  Unfortunately, our current knowledge of structure beyond 8,000~\kms\ is crude and depends on the direction. The reasonably good information to $\sim$16,000~\kms\ in the celestial south provided by the Six Degree Field Redshift Survey \citep{2009MNRAS.399..683J} is insufficiently balanced by our more limited knowledge of the celestial north.  Fortunately, the bulk dipole motion of the volume plays no role in the paths determined in this work as the paths are in the frame of the center of mass of the sample. Furthermore,  because this study has a much larger radius than was used in \cite{2017ApJ...850..207S}, the higher order gravitational influences are greatly diminished.    

We take recourse once again to the 2MASS redshift survey to $K_s=11.75$ \citep{2012ApJS..199...26H} that gives homogeneous representation of the entire sky, save for the Galactic plane cutout, and to the group catalog built from that survey \citep{2015AJ....149..171T}.  We represent overdense structure in a shell $8,000-16,000$~\kms\ in the CMB frame by the 44 clusters in the group catalog with masses greater than $10^{15} \Msun$.  Clearly this description is approximate.  Selection corrections to masses are very large at these redshifts and potentially significant components might be entirely missed.  We work on the hypothesis that the richest clusters delineate regions of the most substantial overdensities.  We hold these elements fixed in co-moving coordinates and following linear theory increase their mass with time to reach masses observed today. 

\section{Numerical Action Method}
\label{sec:nam}

The Numerical Action Method (NAM) for determining the orbits of the components of the large scale structure over cosmological time invokes the variational principle in comoving coordinates as described in \cite{1989ApJ...344L..53P, 1990ApJ...362....1P, 1994ApJ...429...43P}, wherein the physical trajectories are the paths of the extrema and saddle points of the action integral.  All of the derivatives of the action integral with respect to variations of these paths at discrete points in time result in zero.     
Positions trace the center of mass over time of the particles that today are bound to each halo. 

The boundary conditions at the first time step in the action hold that the 3d velocities are low compared to $H_0 d$.  This condition is met because the velocities, as determined by the differences in positions at the first two time steps, following linear theory, are proportional to the gravitational accelerations at the first time step.  The boundary conditions at the final time steps are either the 3d positions or the 2d sky position plus redshifts.

One can start with a random walk of positions backwards from the last step for all particles and take turns improving the orbits by making small positional changes that reduce the sum of the square of all the derivatives (SOS) involving that particle.  As this sum is driven to as close to zero as possible, a process called `path relaxation', the path and the resultant redshift or final position converge to ever more precise values.  The directions and amplitudes of positional changes that relax a path is informed by having the closed form second derivatives of the action integral. Each particle is treated separately in turn while ignoring its effects on the other particles until the next one is to be treated.  Many iterations over all particles are required to reach consistency of the mutual interactions.  

\subsection{Path Calculations}

There is a wide range of precision in our knowledge of the distances to halos in our catalog with the majority having no velocity independent distance determinations.  On the other hand, we typically have redshifts with small uncertainties.  A model can be built by using redshifts to determine distances and, for those entities that have good distance measurements, the model distances can be used to determine a goodness metric. We find paths that have best agreement between model redshift, $cz_m$, and observed redshift, $cz_o$. 

It is imperative that the model positions are not free to drift too far from their measured distances because that would likely imply a gravitational field that is off from reality.  We therefore treat the halos with the highest precision distances differently.    When we have distances with $\mu_e \leq 0.2$ mag, we chose distances that minimize a goodness metric that weighs the ability of the model to reproduce both observed distances and redshifts.  

We use the New NAM (NNAM) algorithms described in \cite{2010arXiv1009.0496P} for speed.  This procedure solves for redshifts given position and then, as desired, we allow individual distances to `drift' to improve the redshift agreement.  When searching for simple redshift agreement, we solve for the paths at two different distances to determine the rate of change of redshift with distance, and calculate the size of the step to take based on the current disagreement.  This procedure usually converges quickly since $cz$ is nearly linear with distance over most of space.  When $cz$ agreement is better than 4~\kms, we stop.  Determination of a best fit (distance, velocity) pair is substantially more time consuming.  Starting at the measured distance, we take a 0.3 Mpc step and grow the step size by 4\% with each step that improves the fit. When a step causes $\chi^2$ to rise, the direction reverses and the step size is halved.  When the step size is reduced to 0.003 Mpc, we stop.  But, because there are triple-valued regions that can trap the final position at the wrong location, we also try two more sequences starting at both $\pm 3\sigma$ locations in distance.  

Two issues complicating solutions are that sometimes NAM finds a very complex path that is not warranted and, at some locations, there may not be a solution.  For complex paths, we go through a procedure discussed in \S~\ref{sec:untie} on untying knots.  For locations with no solution, our procedure is to continue stepping along until a solution can be found and consider this the start of a new sequence.  Throughout, we keep track of the best fit path thus far. 

As in the procedure to find a first solution at initial positions, the distance adjustment procedure must be repeated on all halos until no halo is making significant excursions.  Typically the whole set of orbits reaches a stable solution after a dozen or so iterations.

\subsection{Goodness criteria}

The goodness metric applied to halos is:
\begin{equation}
    \chi^2 = \left(\frac{5\log_{10}({d_m}/{d_o})}{\mu_e}\right)^2 + \left(\frac{cz_m-cz_o}{cz_e}\right)^2
    \label{eq:chi-sq}
\end{equation}
where $d_m$ and $d_o$ are model and observed distances.  The error in redshift is taken to be $cz_e = 60$~\kms\ to account for the possibility that the average velocity of dark matter in a cluster may differ from the average of the cluster galaxies and because the unweighted average that we use (because the data is incomplete) may differ from the mass weighted value appropriate for center of mass motion. 

Since there are multiple plausible paths, care must be taken when comparing \meanchi\ of different scenarios to ensure that the variation is not just due to different paths taken by random chance.  As part of our procedure, we compare individual $\chi = \sqrt{\chi^2}$ values in the most similar previous scenarios to see which $\chi$ values have made big changes for the worse.  Then, by an interactive procedures described in \S~\ref{sec:untie}, attempts are made to see if the better path can be achieved in the present scenario.  

\subsection{Setting up the parameters}

We use 100 time steps to go from $z = 4.0$ to $z = 0$, evenly spaced in the scale factor $a(t)$.  
On the scale of the distances between entities in this model, motions earlier than $z = 4$ are well described by linear perturbation theory, but at later times non-linear interactions are very common.   
We solve for cosmological scenarios keeping $H_0$ in the range from 69 to 77~\kmsMpc.
The Planck Satellite measurement of the acoustic angular scale on the sky sets the coupling between the mean density of matter in the universe and the Hubble Constant: $\Omega_m h^3 = \xi_P = 0.096345$ \citep{2020A&A...641A...6P}.\footnote{$\Omega_m$ is the density of matter, $\rho_m$, with respect to the critical density for a universe closed by matter, $\rho_c$.}
The calculations assume the standard flat $\Lambda$CDM universe (i.e., $\Omega_{\Lambda} + \Omega_m=1$, where $\Omega_{\Lambda} \equiv \frac{c^2\Lambda}{3H_0^2}$ and $\Lambda$ is the cosmological constant) with negligible non-gravitational forces or dynamical friction.
The measured distances and redshifts are never altered.  That is, while $H_0$ is uncertain, we do know distances (with uncertainties) within the constraints of our zero point calibration. 

The $K-$band luminosity to mass mapping for halos is set by Eq.~7 of \cite{2015AJ....149..171T}:  
\begin{equation}
M = \kappa \left(\frac{L_K}{10^{10} \Lsun}\right)^{1.15}10^{10}~\Msun.
\label{eq:mass-to-light}
\end{equation}
The coefficient $\kappa$ in this relation was found to be $43 h$ in that study.  The linear dependence on $H_0$ arises because masses are generally derived from equilibrium dynamics that are independent of time, but proportional to distance, while luminosity is proportional to distance squared.  Hence, mass divided by light is inversely proportional to $H_0$.  In this study, distances are fixed by their velocity independent measurements and masses are not in equilibrium configurations, so a more appropriate dependence of $\kappa$ on $H_0$ needs to be found.   

To account for mass not bound to the regions illuminated by major galaxies, such as in voids, sheets, filaments, low mass halos, or streaming particles, a uniform density interhalo medium, $\Omega_\mathrm{IHM} \equiv \rho_\mathrm{IHM}/\rho_c$, is added to NAM.  With the transformation from physical to comoving coordinates, there is a fictitious outward acceleration term on all particles from empty space \citep[Eq. 20.27]{1993ppc..book.....P} given by the second term in the expression for peculiar gravitational acceleration
\begin{equation}
   \mathbf{g_i} = \frac{G}{a^2}\sum_j m_j\mathbf{\frac{x_j-x_i}{|x_j-x_i|^3}}+ \frac{1}{2}\Omega_m H_0^2 a \mathbf{x_i}.
\end{equation}
To implement the IHM component, the second term is  modified to 
\begin{equation}
    \frac{1}{2}\Bigl(\Omega_m - \frac{a \delta_0 +1}{\delta_0+1}\Omega_\mathrm{IHM}\Bigr) H_0^2 a \mathbf{x_i}.
\end{equation}
This formulation assumes that any present overdensity of the study volume, $\delta_0$, grew to its current amplitude with time dependence a(t). 
The IHM needed in our models must be primarily dark matter since the baryon density is known to be limited to about $0.021 < \Omega_b h^2 < 0.025$ \citep{2014arXiv1412.1408F}.   

\subsection{How to untie a knotted orbit}
\label{sec:untie}
Because we are dealing with many orbits, fairly rare events can be quite numerous. A problem arises when neighboring paths undergo very tight interactions. In reality, such interactions would typically result in a merger after some sloshing back and forth of the orbits.  But, since our calculations do not include dynamical friction, 
mergers do not occur.   Typically, a completed run would have several dozen orbits tied up.
To avoid these unwanted interactions, code was added to help find and interactively untie knots.  
A menu of tools was added to the NAM code to allow the user to select individual paths and choose to 1) set a new distance, 2) find paths backward in time from present position and redshift (`backtrack'), 3) check on the quality of the path ($\chi$ and SOS), 4) relax the path (reduce the SOS) to a prescribed value, 5) step a chosen halo in distance until model redshift agrees with observed, 6) write out path solutions now, in case one is about to make things worse. 

The first step in the procedure to untie an interaction is to `backtrack' at least one of the pair.  A backtrack consists of stepping back in time from the present with a velocity consistent with the observed redshift given a spatial direction.  
The entire $4\pi$ radians of present velocity directions are sampled at $\sim$5\degr\ intervals, and only paths with SOS below a set threshold are displayed.  One of these is chosen, usually the one with lowest SOS if it is not a complicated orbit.  This selection unties the two paths, and sometimes is close to an actual solution, although usually the SOS is still far from its target goal.
Next, one returns to the distance adjustment procedure and allows each to drift until the proper redshift is met.  If the two are still interacting even mildly, the distance adjustments need to be iterated until the paths are stable.

 Additional IDL code was written to locate pairs that are tightly interacting.  The program finds, for each time step, the pairs with the smallest separations. Some pairs come very near to each other but do not suffer a strong interaction, so a list of halos to exclude can be provided.  After untangling a full set of interacting pairs, the knot finder is run again to identify any remaining interactions.
 
 Yet another method to find tight interactions that we used with each scenario is to run an n-body forward integration, with a large number of timesteps, starting with the initial NAM position and velocity, and to compare the final positions with the NAM solution. Tight interactions show up as large disagreements ($> 0.3$ Mpc) in the final positions because the NAM time steps are too large to accurately follow close interactions.
 
\section{Results}
\label{sec:results}

\begin{figure}[ht]
    \centering
    \includegraphics[width=0.47\textwidth]{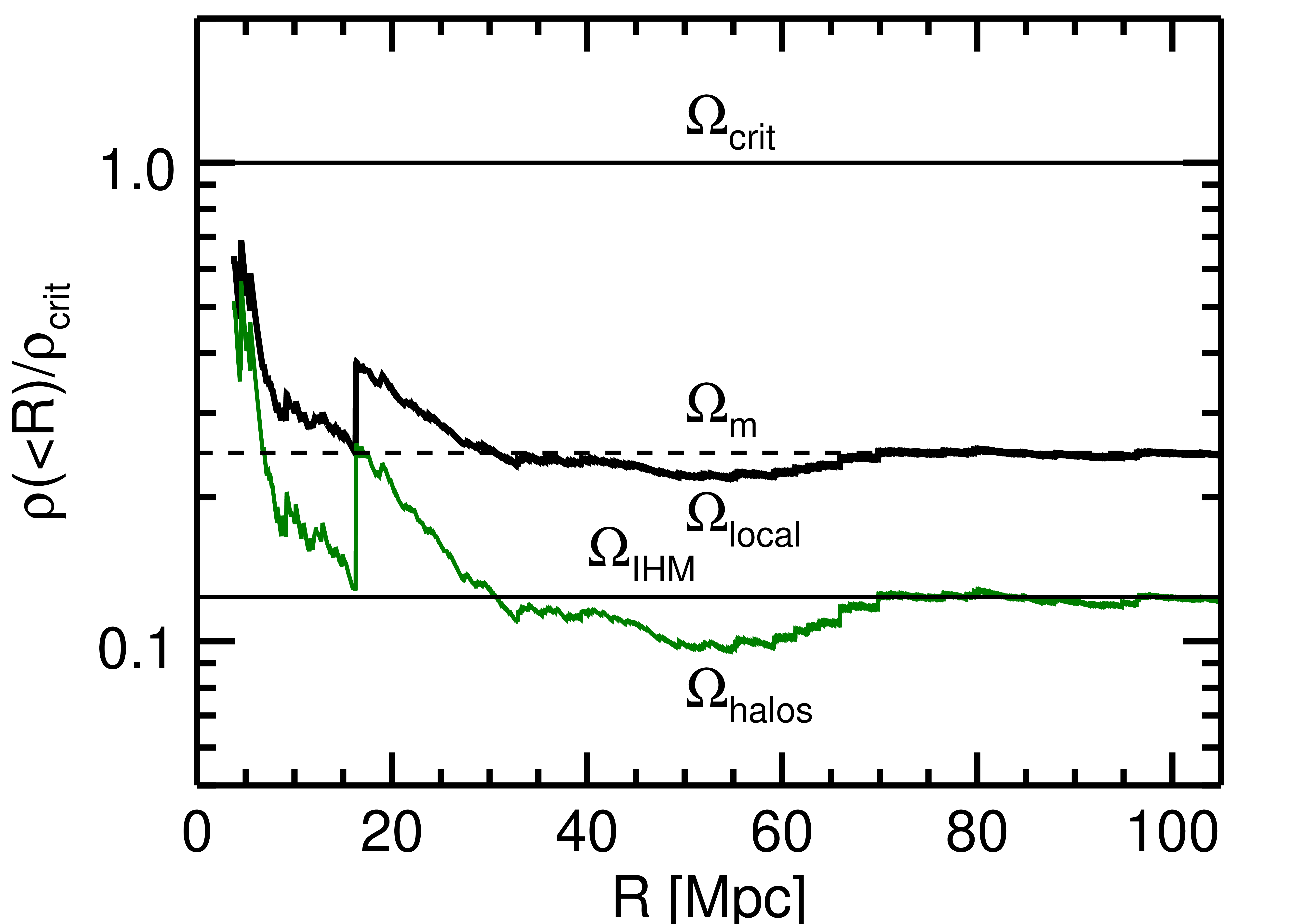}
    \caption{\textbf{Components of density with distance  -}  The sum of the mass density with respect to critical density within radius R from the Local Group in halos, $\Omega_\mathrm{halos}$, plus the IHM, $\Omega_\mathrm{IHM}$, add up to $\Omega_\mathrm{local}$ and is close to $\Omega_m$ at 100 Mpc in model with $H_0=73$, $\Omega_m=0.248$, and $\Omega_\mathrm{IHM}=0.124$.  The jump at 16~Mpc arises from the inclusion of the Virgo Cluster.}
    \label{fig:density}
\end{figure}

We  began our exploration by setting the density of the sample to the mean density of the universe.  
 For each of $H_0 = 69$, 73, and 77 \kmsMpc, three different amounts of galaxy density were tested, $\Omega_\mathrm{halos}/\Omega_m = 0.3$, 0.5, and 0.7.  To attain this, first $\Omega_\mathrm{IHM}$ was set to 0.7, 0.5, and 0.3 of $\Omega_m$.  Then the coefficient $\kappa$ on the luminosity to light mapping was altered until $\delta = \Omega_\mathrm{local}/\Omega_m - 1 \leq \pm 0.01$.  If, at the end of an iteration over all halos, $\delta$ wandered by more than 1\%, $\kappa$ was adjusted.
The resulting distribution of mass density within spheres centered on the Local Group is shown in Figure~\ref{fig:density} for the case of $H_0 = 73$, $\Omega_\mathrm{halos}/\Omega_m = 0.5$.   From 40 to 70 Mpc there is a net underdensity but the density settles to a constant by 70 Mpc.  The other cases gave similar distributions.

For each scenario, we calculated the goodness metric for the precision distance halos ($\mu_e \leq 0.3$ mag) that are presented in Table~\ref{table:runs}: Column 5 is the average value of $\chi$; Column 6 is the median $\chi$.   Columns 7 and 8 have the average and median $\chi_d$ of just the distance term in Eq.~\ref{eq:chi-sq}.  These values indicate, in the best cases, that the NAM distances for the roughly 8,000 halos with poor or no distance measurements have fairly high reliability and should be preferred to their Hubble distances.  Column 9 is the Virgo Cluster $\chi$ to be discussed in \S~\ref{sec:Virgo} and Column 10 has the number of halos with $\chi < 7$ used in the statistics.  

\begin{figure}[ht]
    \centering
    \includegraphics[width=0.47\textwidth]{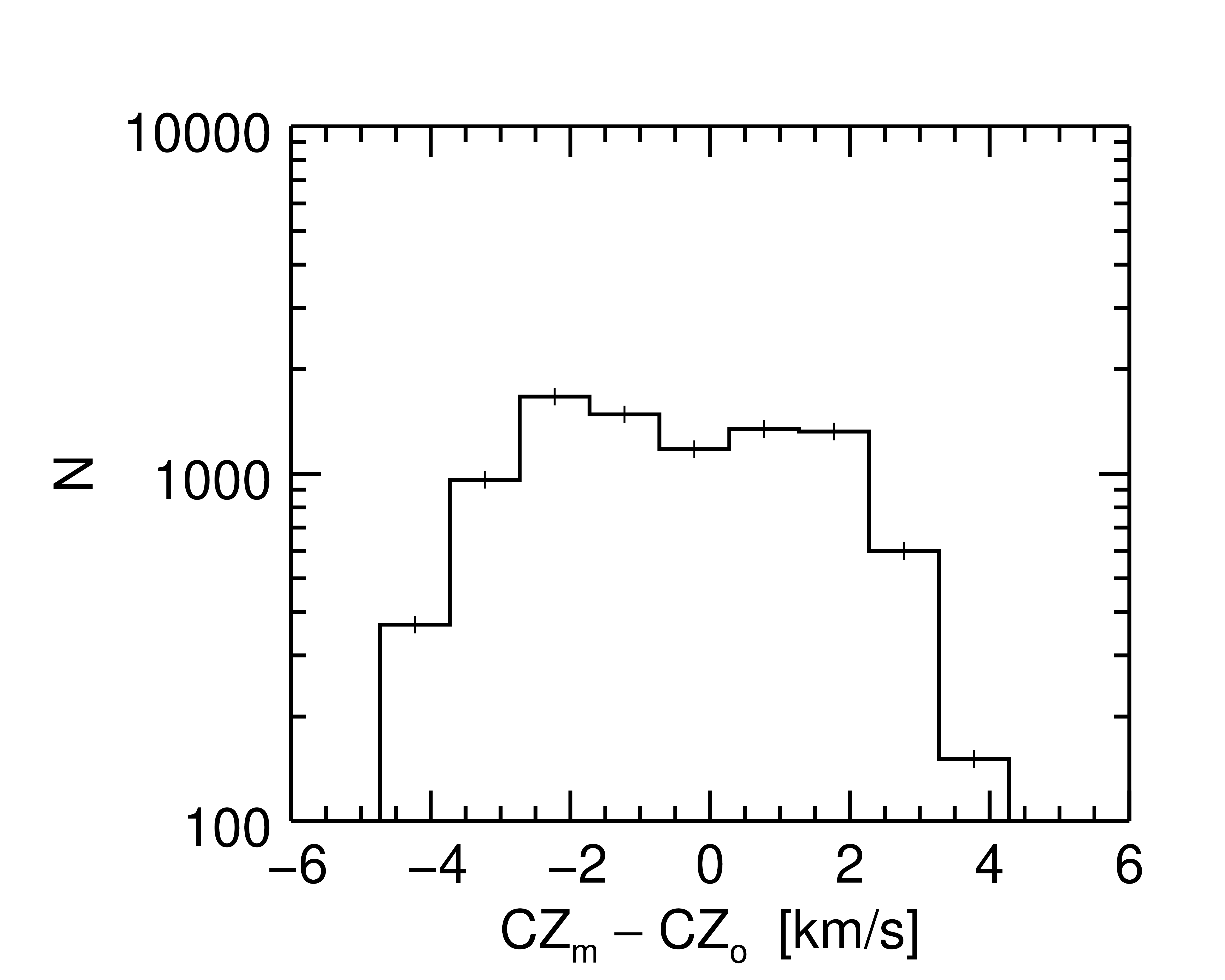}
    \caption{\textbf{Histogram of differences between model and observed redshifts of all halos that have $\mu_e > 0.2$ mag or no $\mu_e$ ($H_0 = 73$ run).}}
    \label{fig:N_delcz}
\end{figure}

\begin{figure}[ht]
    \centering
    \includegraphics[width=0.47\textwidth]{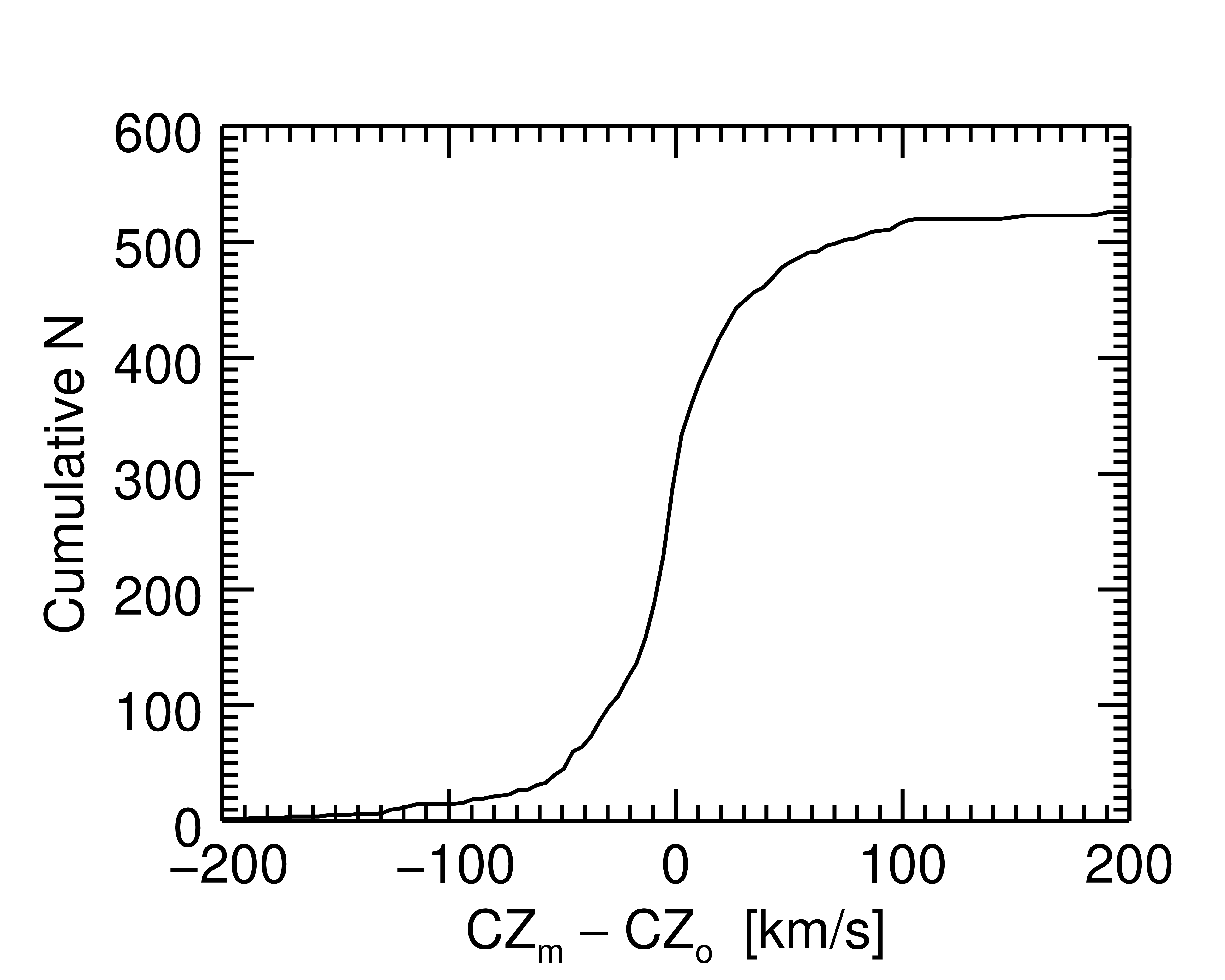}
    \caption{\textbf{Cumulative distribution of differences between model and observed redshifts for halos placed at minima in $\chi$ ($\mu_e \leq 0.2$).}}
    \label{fig:cum_delcz}
\end{figure}

Figure~\ref{fig:N_delcz} shows the distribution of $cz$ agreement for the 9,157 halos either lacking a distance determination or with a distance uncertainty $\mu_e > 0.2$ mag in our fiducial run with $H_0=73$, mass split evenly between halos and the IHM, and $\delta=0$.  We quit seeking better $cz$ agreement when the agreement is $< 4$~\kms. 
Figure~\ref{fig:cum_delcz} shows the cumulative distribution of $cz$ disagreement for the 527 halos with higher precision distances ($\mu_e \leq 0.2$ mag) that have present distances adjusted to minima in $\chi$ (fiducial case).  The vast majority found locations where the $cz$ disagreement is $<70$~\kms.  A few halos fail to find $cz$ agreement near their observed distances, but there are many reasons why that may happen.  
For instance, heavily grouping galaxies can rob a galaxy of a nearby galaxy critical to attaining its $cz_o$ near its true distance.  
Also, in regions around massive structures, there can be multiple solutions with the same radial projected redshifts. 
However, for the  majority of the halos only one reasonable solution is found, probably because most are in the nearly linear perturbation regime.

\begin{figure}[ht]
    \includegraphics[width=0.5\textwidth]{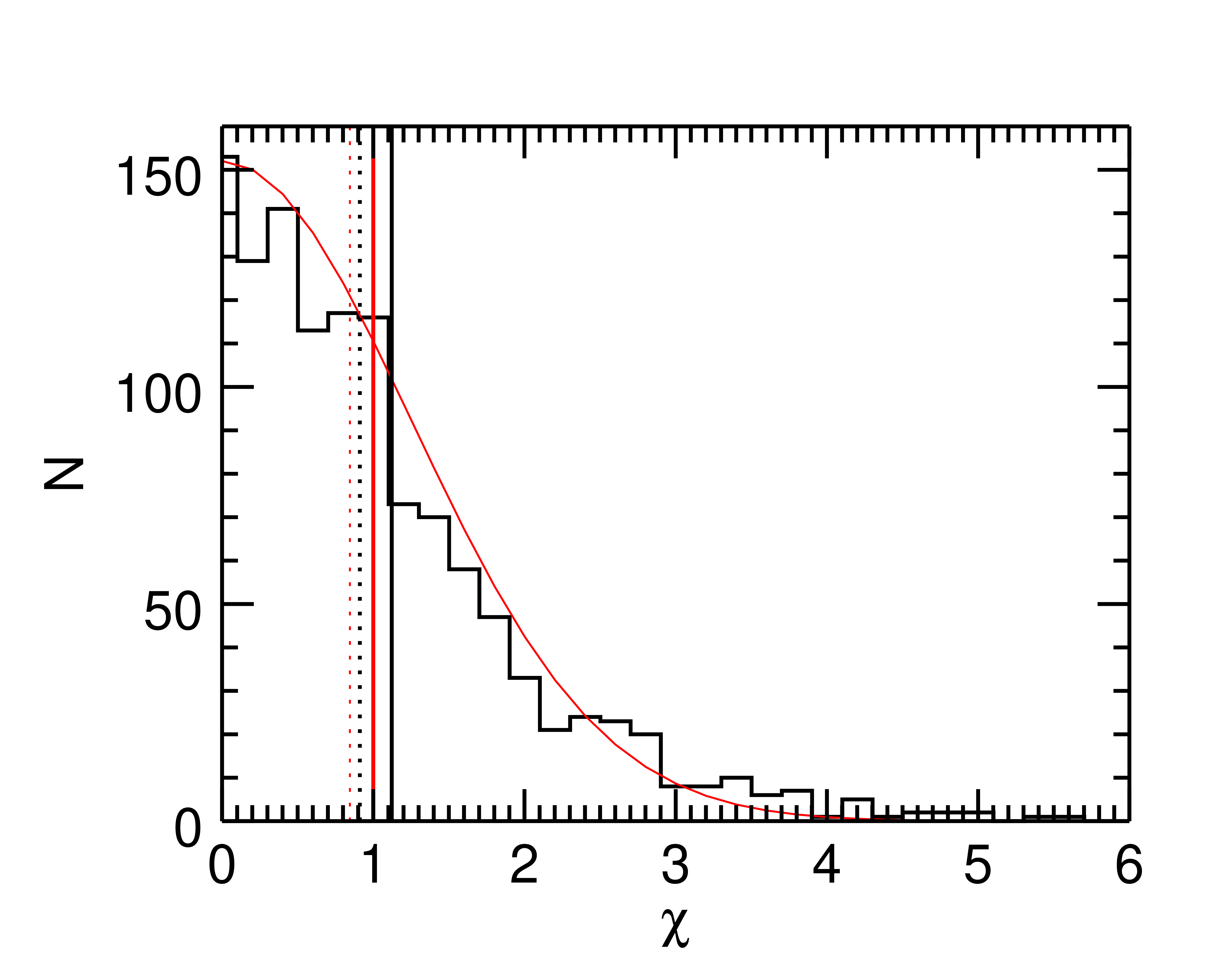}
    \caption{\textbf{Distribution of $\chi$ values} for the $\mu_e \leq 0.3$ halos in the fiducial scenario.  Bin size is 0.2.  Red curve is expected half-normal distribution with 1194 measurements and assuming mean of 1.   The means (black: data; red: half-normal distribution) are solid vertical lines and medians are dotted vertical lines.}
   \label{fig:N_chi}
\end{figure}

\begin{figure}[ht]
\hspace*{-.7cm}
    \includegraphics[width=0.5\textwidth]{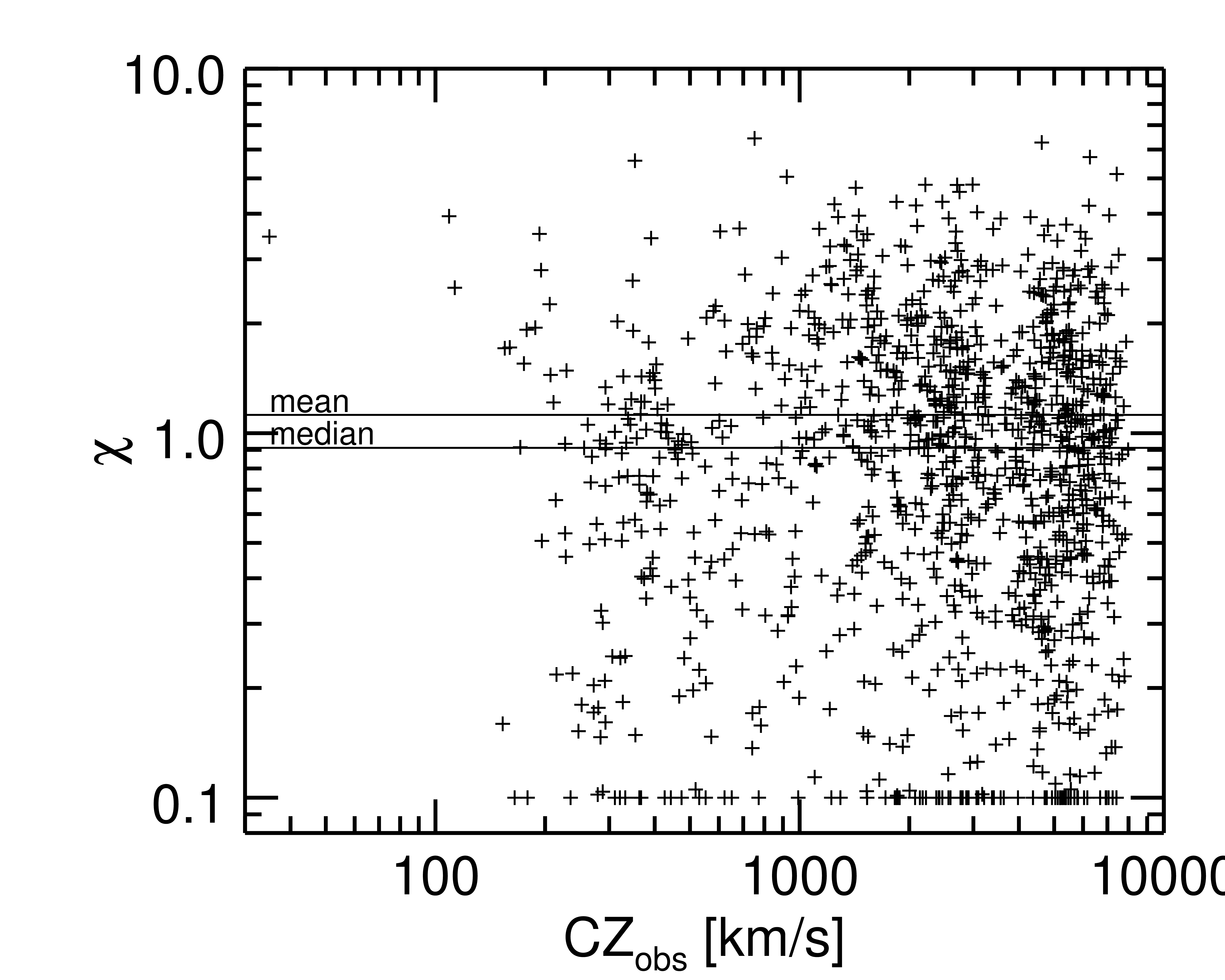}
    \caption{\textbf{Individual $\chi$ values -} for halos with distance errors $\mu_e \leq 0.3$ versus observed redshift for the fiducial scenario.  Values below $\chi = 0.1$ are set to that value to keep them in the plot.  Horizontal lines are placed at the mean and median values.}
    \label{fig:chi_cz}
\end{figure}

 Figure~\ref{fig:N_chi} presents the distribution of $\chi$ for elements with high precision distances for the fiducial case.  The distribution is close to that of a half-normal distribution (red curve) with the same number of measurements and variance. Figure~\ref{fig:chi_cz} presents the individual $\chi$ values as a function of $cz_o$ for that case, with horizontal lines at the mean and median.  
 Values below 0.1 are shown at 0.1 to keep them in the plot.
 
\begin{deluxetable*}{ccc|ccccccc}
\tablewidth{0pt}
\tablecolumns{10}
\tablecaption{Run Parameters and Output\label{table:runs}} \tablehead{
\multicolumn{3}{c|}{Input} & \multicolumn{7}{c}{Output}\\
\dcolhead{H_0} & \dcolhead{\Omega_m} & \dcolheadv{\Omega_\mathrm{IHM}} &  \dcolhead{\delta} &\dcolhead{\langle \chi \rangle} &   \dcolhead{\mathrm{Median}(\chi)} & \dcolhead{\langle \chi_d \rangle} & \dcolhead{\mathrm{Median}(\chi_d)} & \dcolhead{\chi(Virgo)} & \colhead{N}} \startdata 
\multicolumn{10}{c}{$\delta \simeq 0$ runs}\\\hline
69 & 0.2933 &  0.205 & -0.000 &  1.230 &  0.990 &  1.117 &  0.844 &  2.620 &  1195\\
69 & 0.2933 &  0.147 & -0.001 &  1.217 &  0.989 &  1.101 &  0.847 &  3.639 &  1195\\
69 & 0.2933 &  0.088 &  0.000 &  1.315 &  1.071 &  1.200 &  0.965 &  4.304 &  1196\\
71 & 0.2692 &  0.150 &  0.003 &  1.150 &  0.901 &  1.042 &  0.797 &  1.930 &  1195\\
73 & 0.2477 &  0.173 &  0.000 &  1.159 &  0.943 &  1.051 &  0.814 &  1.161 &  1196\\
73\tablenotemark{a} & 0.2477 &  0.124 & -0.001 &  1.120 &  0.902 &  1.006 &  0.772 &  2.640 &  1193\\
73 & 0.2477 &  0.074 & -0.001 &  1.149 &  0.891 &  1.047 &  0.790 &  2.879 &  1195\\
75 & 0.2284 &  0.100 &  0.000 &  1.158 &  0.915 &  1.061 &  0.806 &  2.116 &  1196\\
77 & 0.2195 &  0.148 &  0.001 &  1.305 &  1.054 &  1.178 &  0.941 &  0.248 &  1194\\
77 & 0.2195 &  0.105 &  0.000 &  1.230 &  0.993 &  1.130 &  0.880 &  1.180 &  1196\\
77 & 0.2195 &  0.063 & -0.004 &  1.224 &  1.000 &  1.131 &  0.886 &  1.738 &  1195\\
\cutinhead{$\delta < 0$ runs}
69 & 0.2933 &  0.150 & -0.091 &  1.187 &  0.959 &  1.084 &  0.836 &  2.767 &  1195\\
69 & 0.2933 &  0.120 & -0.181 &  1.156 &  0.936 &  1.058 &  0.802 &  2.479 &  1195\\
69 & 0.2933 &  0.090 & -0.283 &  1.137 &  0.914 &  1.042 &  0.780 &  1.908 &  1195\\
69 & 0.2933 &  0.060 & -0.382 &  1.143 &  0.903 &  1.051 &  0.801 &  1.453 &  1195\\
69 & 0.2933 &  0.045 & -0.429 &  1.152 &  0.920 &  1.055 &  0.800 &  1.445 &  1195\\
69 & 0.2933 &  0.030 & -0.477 &  1.178 &  0.955 &  1.079 &  0.849 &  1.528 &  1195\\
71 & 0.2692 &  0.122 & -0.099 &  1.126 &  0.890 &  1.031 &  0.782 &  2.354 &  1195\\
71 & 0.2692 &  0.108 & -0.151 &  1.120 &  0.887 &  1.024 &  0.789 &  2.207 &  1195\\
71 & 0.2692 &  0.090 & -0.215 &  1.121 &  0.904 &  1.021 &  0.780 &  2.043 &  1195\\
\cutinhead{$\delta > 0$ runs}
75 & 0.2284 &  0.112 &  0.051 &  1.150 &  0.891 &  1.059 &  0.809 &  2.174 &  1196\\
75 & 0.2284 &  0.127 &  0.098 &  1.141 &  0.902 &  1.050 &  0.797 &  2.313 &  1196\\
75 & 0.2284 &  0.139 &  0.147 &  1.137 &  0.898 &  1.044 &  0.799 &  2.121 &  1196\\
75 & 0.2284 &  0.150 &  0.195 &  1.146 &  0.898 &  1.053 &  0.807 &  2.823 &  1196\\
77 & 0.2195 &  0.136 &  0.237 &  1.173 &  0.924 &  1.078 &  0.810 &  1.993 &  1195\\
77 & 0.2195 &  0.150 &  0.300 &  1.167 &  0.927 &  1.065 &  0.809 &  2.273 &  1196\\
77 & 0.2195 &  0.161 &  0.351 &  1.165 &  0.924 &  1.065 &  0.812 &  2.294 &  1196\\
77 & 0.2195 &  0.172 &  0.398 &  1.164 &  0.922 &  1.066 &  0.809 &  2.446 &  1196\\
77 & 0.2195 &  0.190 &  0.484 &  1.181 &  0.932 &  1.080 &  0.835 &  3.327 &  1195\\
\enddata
\tablenotetext{a}{fiducial scenario}
\end{deluxetable*}

\begin{deluxetable*}{rrlrrrrrrr|rrrrrrrr}
\rotate
\tablewidth{0pc}
\tabletypesize{\scriptsize}
\tablecolumns{18}
\tablecaption{Input/Output Halo Catalog for $H_0 = 73, \delta=0, \Omega_\mathrm{halos} = 0.5 \Omega_m$ Scenario\label{table:results}}
\tablehead{
\multicolumn{10}{c|}{Input} & \multicolumn{8}{c}{Output}\\
\colhead{1PGC} & \colhead{Halo} & \colhead{Name}
 & \colhead{SGL} & \colhead{SGB} &\dcolhead{N_g} 
& \dcolhead{o_{d_o}} & \dcolhead{d_o} & \dcolhead{\mu_e} & \dcolheadv{cz_o} &
\dcolhead{M_{K_s}} & \dcolhead{d_{m-o}} & \dcolhead{cz_{m-o}} & \colhead{Log(M)} & \dcolhead{V_{sgx}} & \dcolhead{V_{sgy}}
& \dcolhead{V_{sgz}} & \dcolhead{\chi} \\
\colhead{} & \colhead{} & \colhead{}
 & \colhead{deg} & \colhead{deg} &\colhead{} & \colhead{} 
& \colhead{Mpc} & \colhead{mag} & \dcolheadv{\mathrm{km~ s^{-1}}} & \colhead{Mag} & 
\colhead{Mpc} & \colhead{\kms} & \dcolhead{\Msun} & \colhead{\kms} & \colhead{\kms}
& \colhead{\kms} & \colhead{}} 
\colnumbers
\startdata
  44715 & 100001 & Coma Cl      &  89.6226 &   8.1461 & 136 & 106 &  96.4 & 0.05 & 6926 & -30.199 & -0.0 &    1 &  15.246 &  -162 &  183 & -217 & 0.03 \\
  57612 & 200002 & ESO137-006   & 188.2039 &   6.9076 & 167 &   0 &  65.3 & 0.00 & 4767 & -29.964 &  0.6 &   -3 &  15.147 &  -105 &  142 &  -95 & 0.00 \\
  16124 &      0 &              &   2.6934 & -23.1887 &   6 &   0 &  93.9 & 0.00 & 6854 & -29.789 &  1.0 &   -1 &  15.067 &  -141 & -149 &   25 & 0.00 \\
  12429 & 200001 & Perseus Cl   & 347.7159 & -14.0594 & 180 &  44 &  67.3 & 0.06 & 5546 & -29.756 &  2.5 &  -30 &  15.079 &   205 & -115 &  -47 & 1.43 \\
  70042 & 200450 & NGC7426      & 330.2136 &  33.2224 &   6 &   1 & 141.3 & 0.40 & 5812 & -29.725 &-69.1 &    0 &  14.357 &   189 & -118 &  -51 & 0.00 \\
 166288 &      0 &              & 174.1307 &  -9.8597 &  38 &   0 &  74.9 & 0.00 & 5465 & -29.289 &  4.8 &    1 &  14.889 &    89 &  194 &  144 & 0.00 \\
  36487 & 100005 & Leo Cl       &  92.0255 & -10.4950 &  61 &  28 &  88.3 & 0.06 & 6458 & -29.189 & -0.3 &    1 &  14.778 &   -76 &  408 &  193 & 0.12 \\
  63247 & 200228 & ESO460-004   & 216.2660 &  44.1999 &  11 &   2 & 137.4 & 0.35 & 7282 & -28.974 &-40.5 &    2 &  14.333 &  -109 & -100 &  -96 & 0.00 \\
   4149 & 200018 & Abell 2877   & 251.7068 & -13.6204 &  32 &  25 & 102.8 & 0.09 & 6876 & -28.871 &-10.0 &   24 &  14.533 &   114 &   10 &  -48 & 2.49 \\
   9201 & 200013 & Abell 347    & 343.2559 &  -5.2785 &  54 &  13 &  78.0 & 0.12 & 5940 & -28.801 &  2.1 &   -5 &  14.628 &  -189 &   64 & -483 & 0.48 \\
2801981 &      0 &              & 184.6380 &   6.3997 &  17 &   0 &  67.1 & 0.00 & 4896 & -28.639 &  9.5 &   -2 &  14.660 &   543 &   76 &  -78 & 0.00 \\
  11188 & 200012 & Abell 400    & 312.1102 & -27.2311 &  44 &  33 &  92.9 & 0.07 & 7366 & -28.626 &  3.7 &  -19 &  14.561 &   164 &  -25 &   48 & 1.26 \\
   6962 & 200003 & Abell 262    & 334.9977 &  -2.0249 &  63 &  37 &  66.7 & 0.06 & 5102 & -28.621 & -5.3 &   91 &  14.437 &   371 & -189 & -216 & 3.37 \\
\multicolumn{11}{c}{ ... data continues ...} \\
\enddata
\tablecomments{Column descriptions: (1) PGC ID of brightest halo member; (2) Halo ID; (3) Name of brightest halo member; (4) Supergalactic longitude; (5) Supergalactic latitude; (6) Number of galaxy members brighter than $K_s = 11.75$ mag; (7) Number of distance measurements; (8) Observed distance or Hubble distance if no measures; (9) Distance modulus error; (10) Observed radial velocity in Local Group frame; (11) Output absolute $K_s$-band magnitude ($M_{K_s}=3.27$ is assumed for the Sun); (12) Model minus observed distance; (13) Model minus observed radial velocity; (14) Model Log$_{10}$ of Mass; (15,16,17) Supergalactic vx,vy,vz in frame of sample (18) Goodness measure. }
   \tablecomments{Table \ref{table:results} is published in its entirety in the machine-readable format. A portion is shown here for guidance regarding its form and content.}
\end{deluxetable*}

Table~\ref{table:results}
provides input and output data on each of the individual halos of the fiducial $H_0 = 73, \Omega_\mathrm{halos}/\Omega_m = 0.5$ calculation with local density at the mean density.  Only a few rows are presented here, but the entire machine readable table is available at the online journals, and it has been installed into the Extragalactic Distances Database (https://edd.ifa.hawaii.edu).

\begin{figure}[ht]
\hspace*{-.4cm}
    \includegraphics[width=0.5\textwidth]{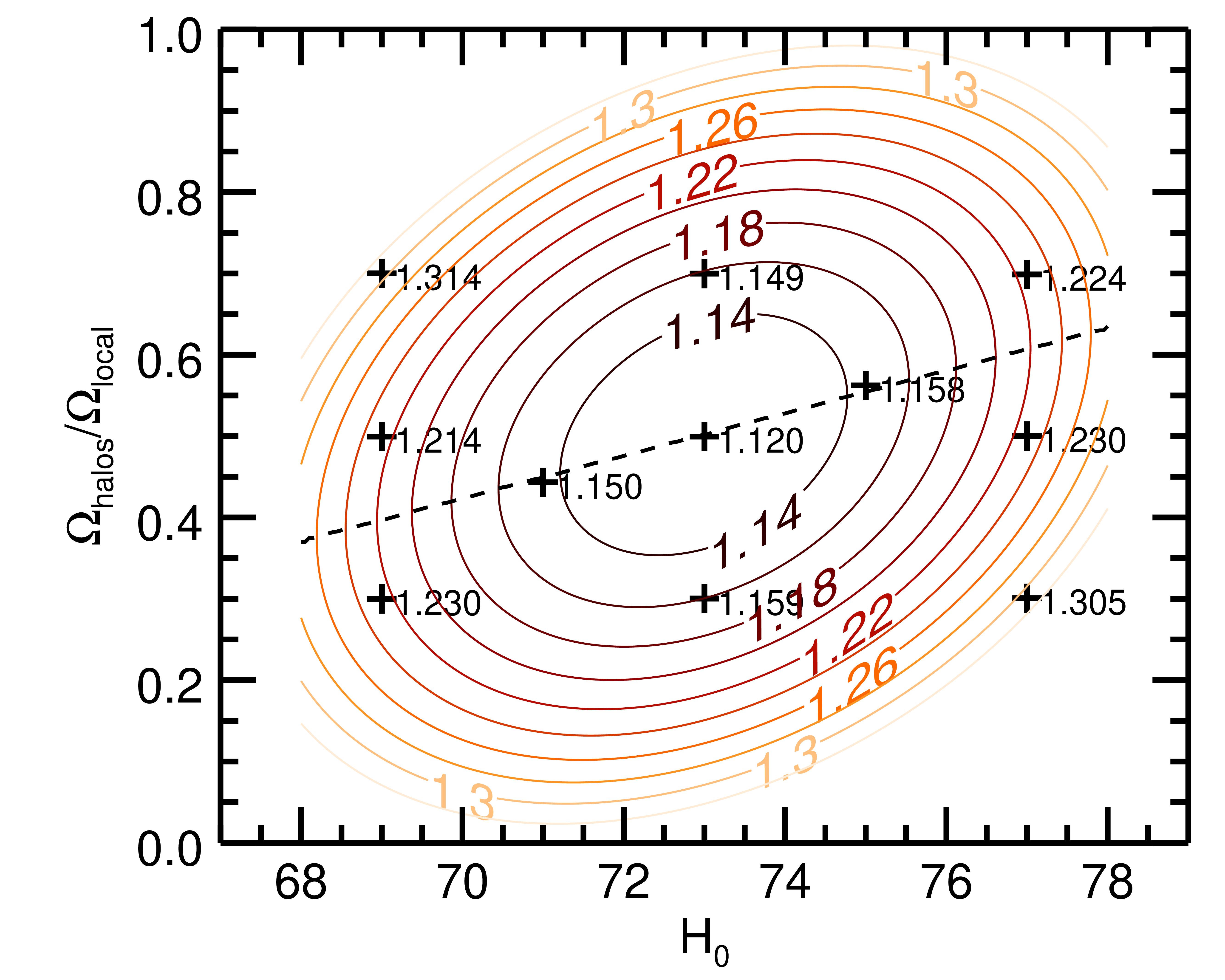}
    \caption{\textbf{\meanchi\  as a function of $\mathbf{H_0}$ and fraction of mass density in halos.}
    Goodness metrics, mean $\chi$ values, are shown next to plus marks for 3 runs at $H_0$= 69, 73, and 77 \kmsMpc\  ($\Omega_m h^3 =\xi_P$) and mass-to-light ratios varied to obtain $\Omega_\mathrm{halos}/\Omega_m$ = 0.3, 0.5, and 0.7.  Additionally, there are single runs at $H_0 = 71$ and 75. Adding the IHM density brings the total density to the global mean.  The \meanchi\ values are fit with a 2d second order polynomial, shown as contours. The minimum $\langle \chi \rangle$ of the fit occurs at $H_0$=72.95.  The minima at each $H_0$ is shown as a dashed line.
    }
    \label{fig:Ogal_ratio}
\end{figure}

\begin{figure}[ht]
\centering
\begin{minipage}[h]{\linewidth}
\includegraphics[width=\linewidth]{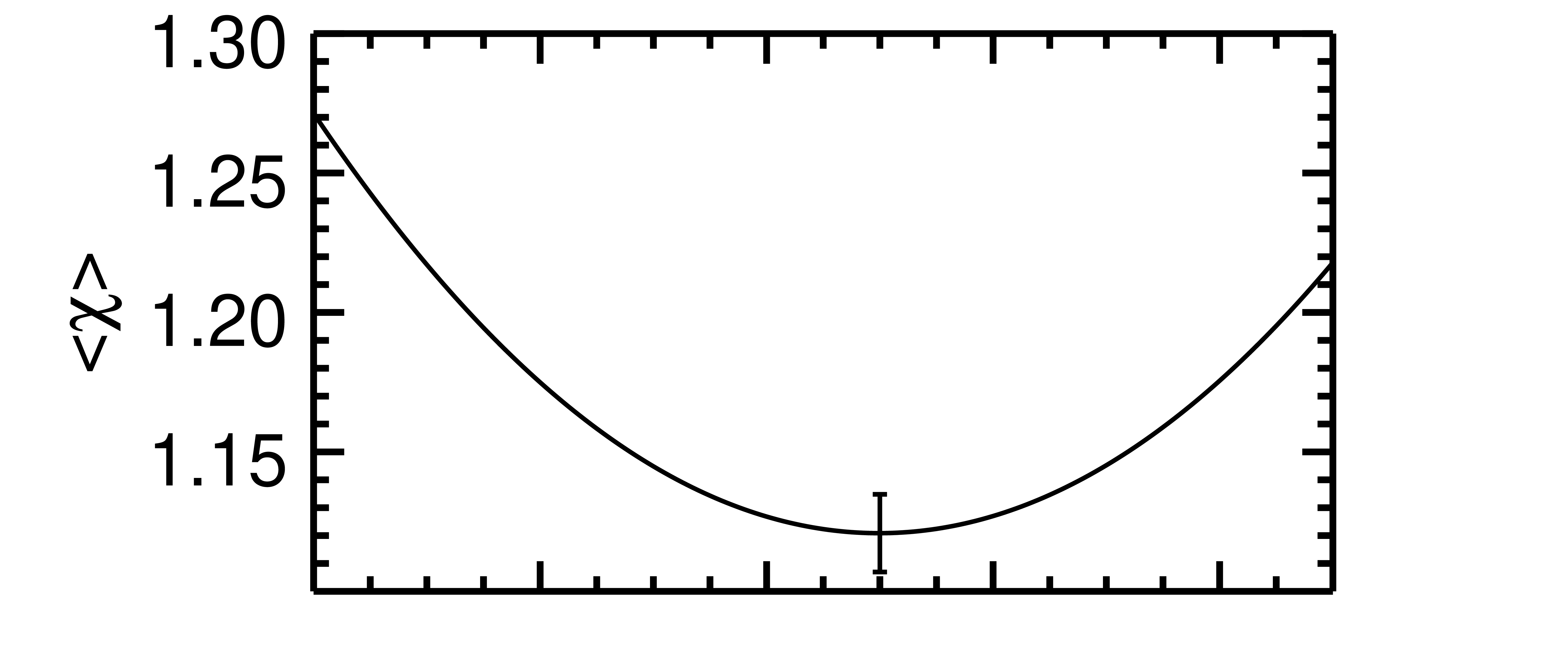}
\end{minipage}

   \vspace{-16 pt}
    
\begin{minipage}[h]{\linewidth}
\includegraphics[width=\linewidth]{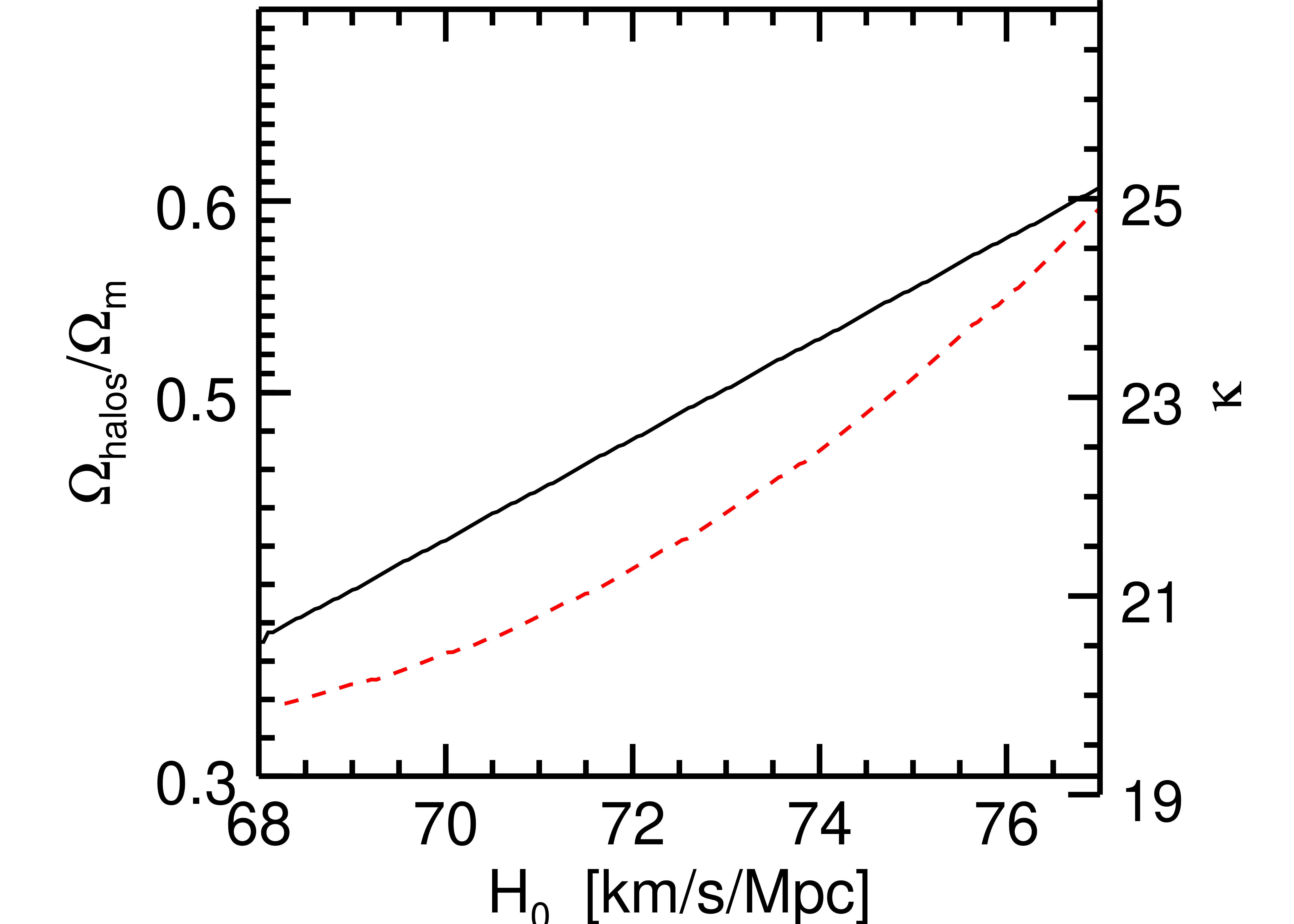}
\end{minipage}
\caption{\textbf{Parameters along the minima $\chi$ -}
 The top curve  traces along the minima \meanchi\ at each $H_0$ value of the fit to $\delta=0$ solutions (previous figure). Bottom, the  $\Omega_\mathrm{halos}/\Omega_m$ along the trace is shown by solid black line.  The red dash line provides the approximate value of coefficient $\kappa$ in the luminosity to mass mapping (use axis on the right).
 \label{fig:alongmin} }
\end{figure}

Figure~\ref{fig:Ogal_ratio} presents the goodness metric, \meanchi\ for halos with $\mu_e \leq 0.3$, as functions of both $\Omega_\mathrm{halos}/\Omega_m$ and $H_0$.  The values for 11 model runs at $\delta=0$ are shown at the plus signs.  Halos with $\chi > 7$ were omitted from the mean (usually just an object or two).
We also omitted halos with observed distances $d_o < 3$ Mpc because they are too sensitive to small changes in the model velocity of the Local Group.  A second order polynomial in two dimensions is fit to the data and is shown as contours.  The minima at each $H_0$ were found and they occur along the dashed line.  The overall minimum of the fit is very close to the fiducial case.
Minimum \meanchi\ versus $H_0$ is shown in the top section of Fig.~\ref{fig:alongmin}.  The typical 1$\sigma$ error bar is shown at its minimum at $H_0=73.0$.   In addition, we have transformed the location of the minima into $\kappa$, the coefficient in the mass-to-light ratio formula, seen as the red dashed line and using the right hand axis. A value of $\kappa = M/L_K = 21.9~\Msun/\Lsun$ is preferred at $H_0=73$.  The greater than  linear increase of mass with $H_0$ is attributed to the decrease in age of the universe and the increase in $\Omega_{\Lambda}$.

\subsection{$\delta \neq 0$}
\citet{2016AJ....152...50T} demonstrated (their Figure 21) that, with the {\it Cosmicflows-3} distances used here, global outflow from the region within 8,000~\kms\ is anticipated for values of $H_0$ less than $\sim$74~\kmsMpc\ and infall for larger $H_0$ values.  Demonstrably, within a radius of 30~Mpc we live in a local overdensity (see Figure~\ref{fig:density}). On the full scale of this study, there have been suggestions that we live in an underdense region  \citep{2013ApJ...775...62K, 2019MNRAS.490.4715S, 2020A&A...633A..19B}.

\begin{figure}[ht]
    \centering
    \includegraphics[width=\linewidth]{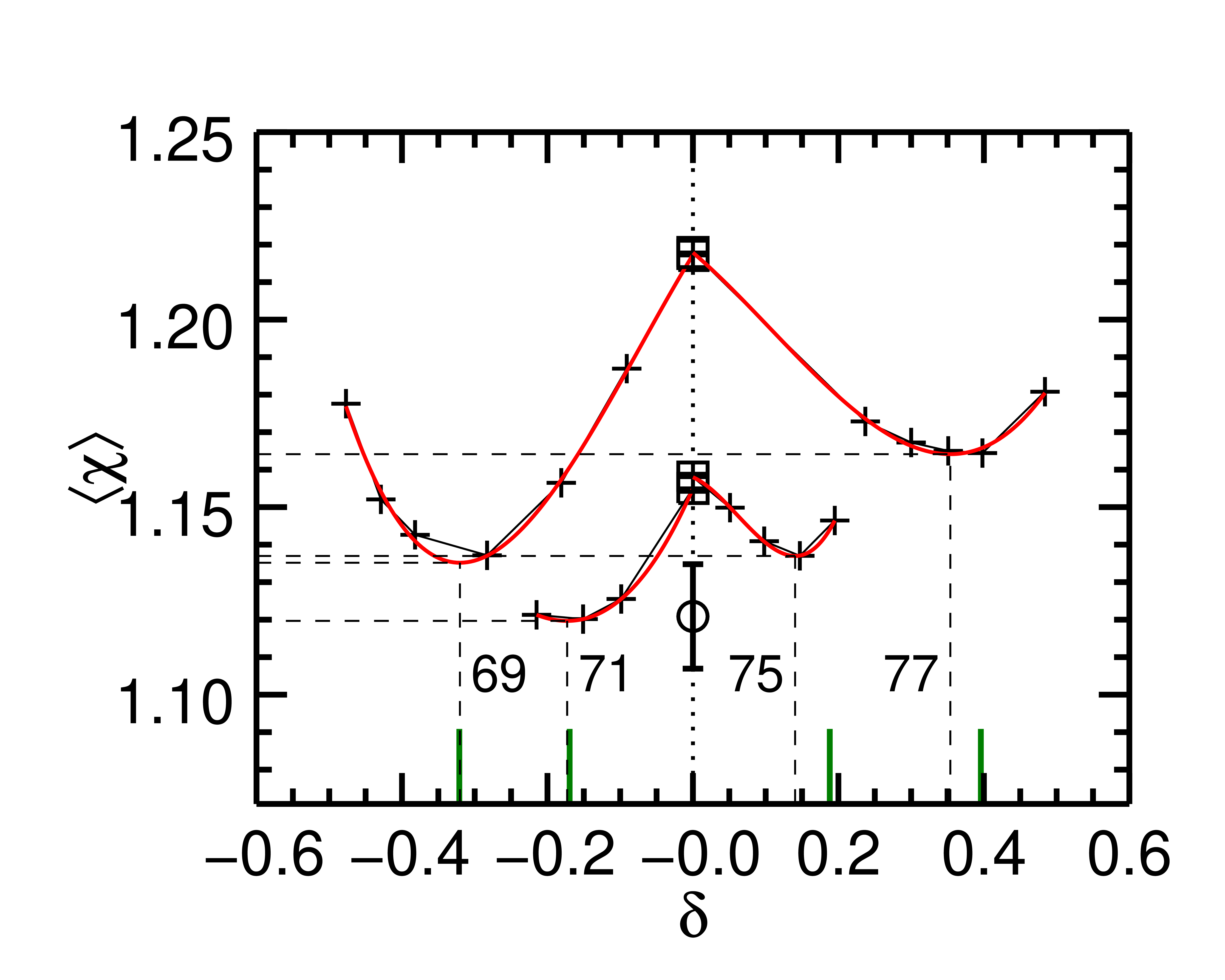}
    \caption{\textbf{Dependence of \meanchi\  on over/underdensity at select values of $\mathbf{H_0}$ -} A series of runs with different $\Omega_\mathrm{IHM}$ at each $H_0$ value were fitted to find where the minimum \meanchi\  occurs as a function of local (100 Mpc) overdensity in the models.  The red lines are 3rd order polynomial fits.  Vertical dashed lines drop from the minima of each fit and the values of $H_0$ are given next to this line.  The horizontal dashed lines are placed at the minima of the fits. The circle is placed at the minimum of the solutions with $\delta = 0$ in the previous 2 figures.  The green ticks at the bottom are analytical predictions from integration of the Friedmann equation (\S~\ref{sec:deltas}).
    \label{fig:delta_chi}}
\end{figure}

\begin{figure}[ht]
    \centering
    \begin{minipage}[h]{\linewidth}
    \includegraphics[width=.9\linewidth]{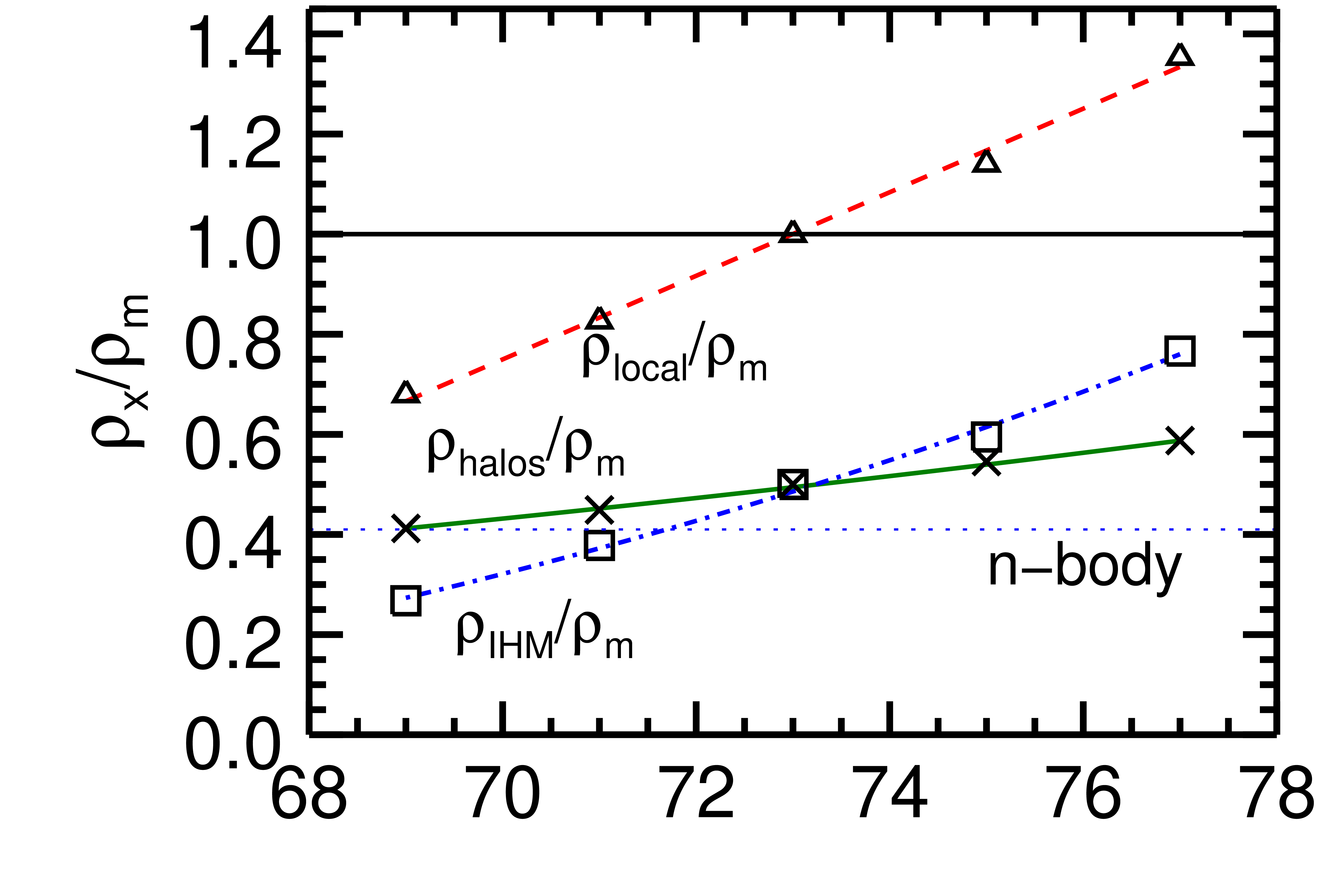}
    \end{minipage}
    
    \vspace{-29 pt}

\begin{minipage}[h]{\linewidth}
\includegraphics[width=.9\linewidth]{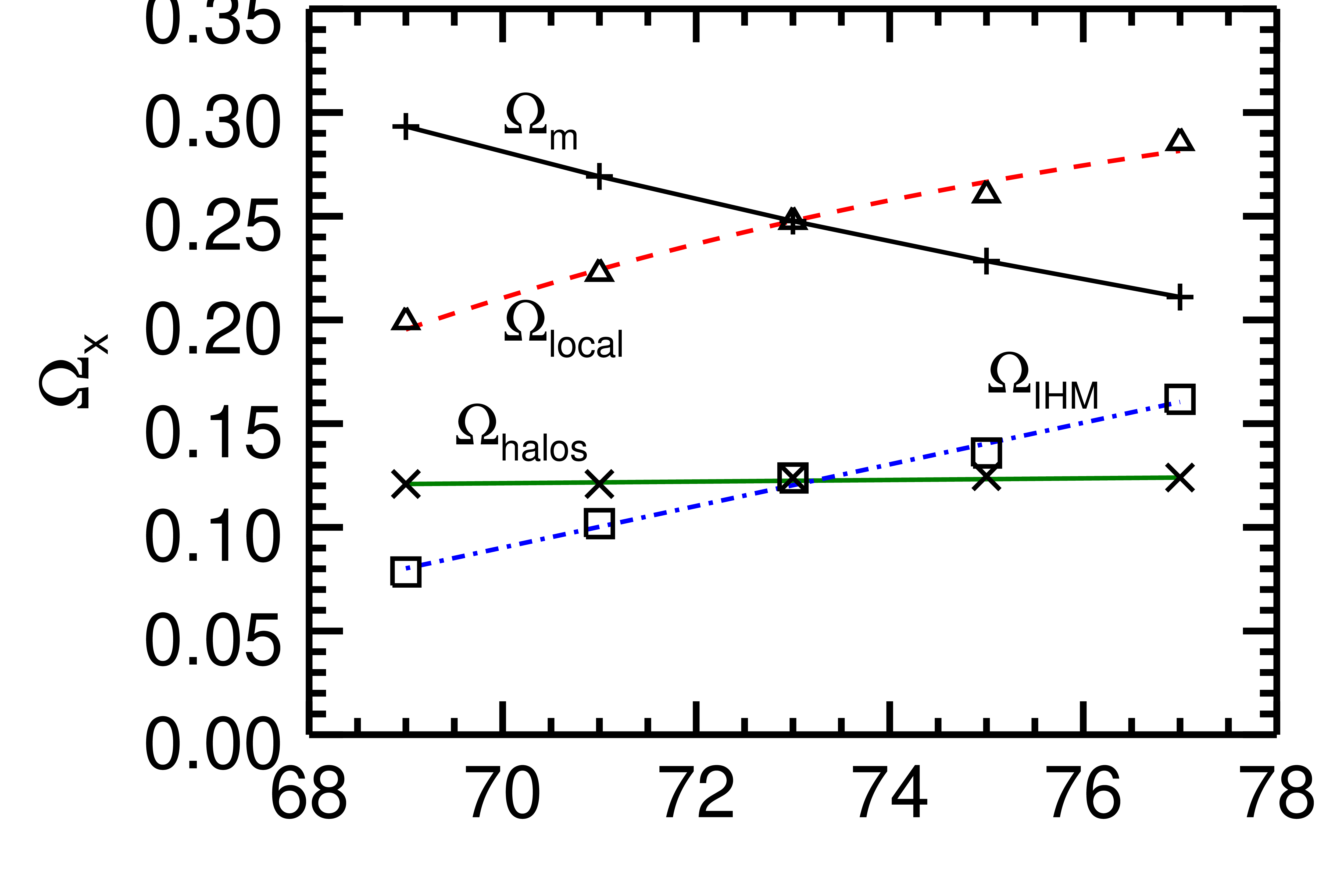}
\end{minipage}
    \caption{\textbf{Mass density components along valley of lowest \meanchi},  This diagram delineates the valley where \meanchi\  is a minimum as the IHM density is varied for each $H_0$ (see previous figure).  (Top) The crosses reflect the resulting fraction of density in halos,  and the green line through them is  the fit, Eq.~\ref{eq:halos}.  The squares give the fraction of density attributed to the IHM, and the dash-dotted blue line uses the fit, Eq.~\ref{eq:OIHM}.  The triangles show the sum of the two mass components.  The fiducial $H_0=73$ run with $\delta=0$ is included and appears to lie along the \meanchi\ valley. 
    Bottom plots are the same as the top but densities are divided by the critical densities rather than the global mean densities.  The thin dotted horizontal line in the top figure sits at the level of  $\Omega_\mathrm{IHM}/\Omega_m$ expected from n-body simulations (discussed in \S~\ref{sec:simulations}).
    \label{fig:Ox_H}}
\end{figure}

The \meanchi\  arising from assuming a range of $\Omega_\mathrm{IHM}$ values in NAM runs, relaxing the requirement of no overall over/under density in the study volume, are shown  at $H_0$ values of 69, 71, 75 and 77 in Fig.~\ref{fig:delta_chi}.
These strings of values are fit with third-order polynomials.  The vertical dashed lines are placed at the fractional overdensity value where \meanchi\  reaches its minima.
A circle is placed in Figure~\ref{fig:delta_chi}  at the minimum value of the $\delta = 0$ runs in Figure~\ref{fig:Ogal_ratio} that occurs at $H_0=73$, and the 1$\sigma$ errorbar is added.  

It is to be noted that solutions with acceptable values of \meanchi\  can be found from $H_0 = 69$ to 75.  
Increasing $\Omega_\mathrm{IHM}$ to the optimal value at $H_0=75$ results in a \meanchi\  that is only about $1\sigma$ more than our best case. 
The values of $\rho_\mathrm{IHM}/\rho_m$, $\rho_\mathrm{halos}/\rho_m$, and  $\rho_\mathrm{local}/\rho_m$ at the minima in \meanchi\  are shown in Figure~\ref{fig:Ox_H}. 
For $H_0=69$, a very low value of $\Omega_\mathrm{IHM}$ is required, implying a best fit underdensity of $\sim$32\%. 
These optimal scenarios, summarized in Figure~\ref{fig:Ox_H}, are described by the relationship between $H_0$ and $\delta$:

\begin{equation}
    \frac{H_0}{H_{\delta=0}} = 1 + 0.165 \delta
    \label{eq:delta}
\end{equation}
where $H_{\delta=0}$ = 73.0.  
This result can be compared with the related problem of the fractional change in a galaxy's expansion rate $H_i$ expected in linear perturbation theory \citep[p. 116]{1991MNRAS.251..128L, 1993ppc..book.....P}, 
\begin{equation}
\frac{H_0 - H_i}{H_0} = - \frac{v_{pec}}{H_0d} = \frac{\Omega_m^{0.55}}{3}\delta
\end{equation}
The linear theory coefficient on $\delta$ for $\Omega_m = 0.248$ is 0.155.

The values of $\Omega_\mathrm{halos}$ can be thought of as an input parameter in that the mass-to-light relation $\kappa$ was varied in accordance with the fit in Fig~\ref{fig:alongmin}. It is nearly constant and fit by:
\begin{equation}
  \Omega_\mathrm{halos} = 0.122 +  3.9 \times 10^{-4} (H_0 - 73.0).
   \label{eq:halos}
\end{equation}
An expression fitting the run of $\Omega_\mathrm{IHM}$ is:
\begin{equation}
    \Omega_\mathrm{IHM}  = 0.612 \left(\frac{H_0}{H_{limit}} -1\right)
    \label{eq:OIHM}
\end{equation}
where $H_{limit}=61.2$ is the $H_0$ value below which the density of the IHM at the bottom of the \meanchi\ valley becomes negative.

\subsection{Top hat solutions to expansion rate discrepancy}
\label{sec:deltas}
Since the overall minimum \meanchi\ for $\delta=0$ cases occurs very close to the fiducial $H_0 = 73$, we can expect that in each flat universe with ($H_0$, $\Omega_m$) pairings that the optimal density for the local region will be close to one that obtains a local expansion of 73 \kmsMpc.  
Thus, the density required to embed inside a flat universe of $(h, \Omega_m,\Omega_\Lambda) = (h,\xi_P/h^3,1-\xi_P/h^3)$ a homogeneous subregion with  $\Omega_m=\Omega_m^{73}$ with expansion rate of $h=0.73$ provides additional estimates of the parameters of the valley of minimum \meanchi\ and a check on the NAM analysis.  
By setting the age and the cosmological constant, $\Lambda \propto \Omega_{\Lambda}h^2$, of the embedded scenarios to be the same as in the exterior, we can solve for the desired subregion density $\Omega_m^{73}$ in:
\begin{equation}
   t_0(h,\Omega_m,\Omega_{\Lambda}) = t_0(0.73,\Omega_m^{73},\Omega_{\Lambda}h^2/0.73^2).
\end{equation}
The subregion scenario has no closed form formula for $t_0$, but it can be found by integrating the solution for the Friedmann equation:

\begin{equation}
    H_0 t_0 = \int_0^1 \frac{da}
    {\left[\Omega_m a^{-1} + \Omega_{\Lambda} a^2 + (1 - \Omega_m - \Omega_{\Lambda})
    \right]^\frac{1}{2} 
    }
\end{equation}

To convert $\Omega_m^{73}$ to an overdensity, we note that mass density varies $\propto \Omega_m h^2$, so the overdensity is:
\begin{equation}
\delta = \frac{0.73^2 \Omega_m^{73}h}{\xi_P}-1
\end{equation}
For $H_0= (67,69,71,75,77)$ one finds the estimates for $\delta$\ are
(-0.456, -0.321, -0.169, 0.188, 0.396) as shown in Figure~\ref{fig:delta_chi} as green ticks at the bottom.  For negative $\delta$, the NAM calculated best $\delta$s are in very good agreement with these values, but at the higher values there is substantial disagreement.   This discrepancy from the top hat model is to be expected, because adding density to regions in the non-linear regime will alter velocities much more than in low density regions.  This limits how much IHM can be added before some regions badly overreact.  


\subsection{Matter outside of halos in simulations}
\label{sec:simulations}

N-body simulations can predict the amount of matter in and not in luminous independent halos.  We can check if our results are consistent with simulations, and whether simulations favor a portion of the range of our results.
There are numerous papers on the mass function of halos found in simulations \citep{1999MNRAS.308..119S,2001MNRAS.321..372J,2006ApJ...646..881W,2008ApJ...688..709T,2013MNRAS.433.1230W,2020ApJ...903...87D}.   The study of observed halo masses by \citet{2017ApJ...843...16K} suggests that the luminous halos in our catalog can be associated with almost all of the mass in halos above $2 \times 10^{10} \Msun$.  
\begin{figure}[ht]
    \centering
    \includegraphics[width=.5\textwidth]{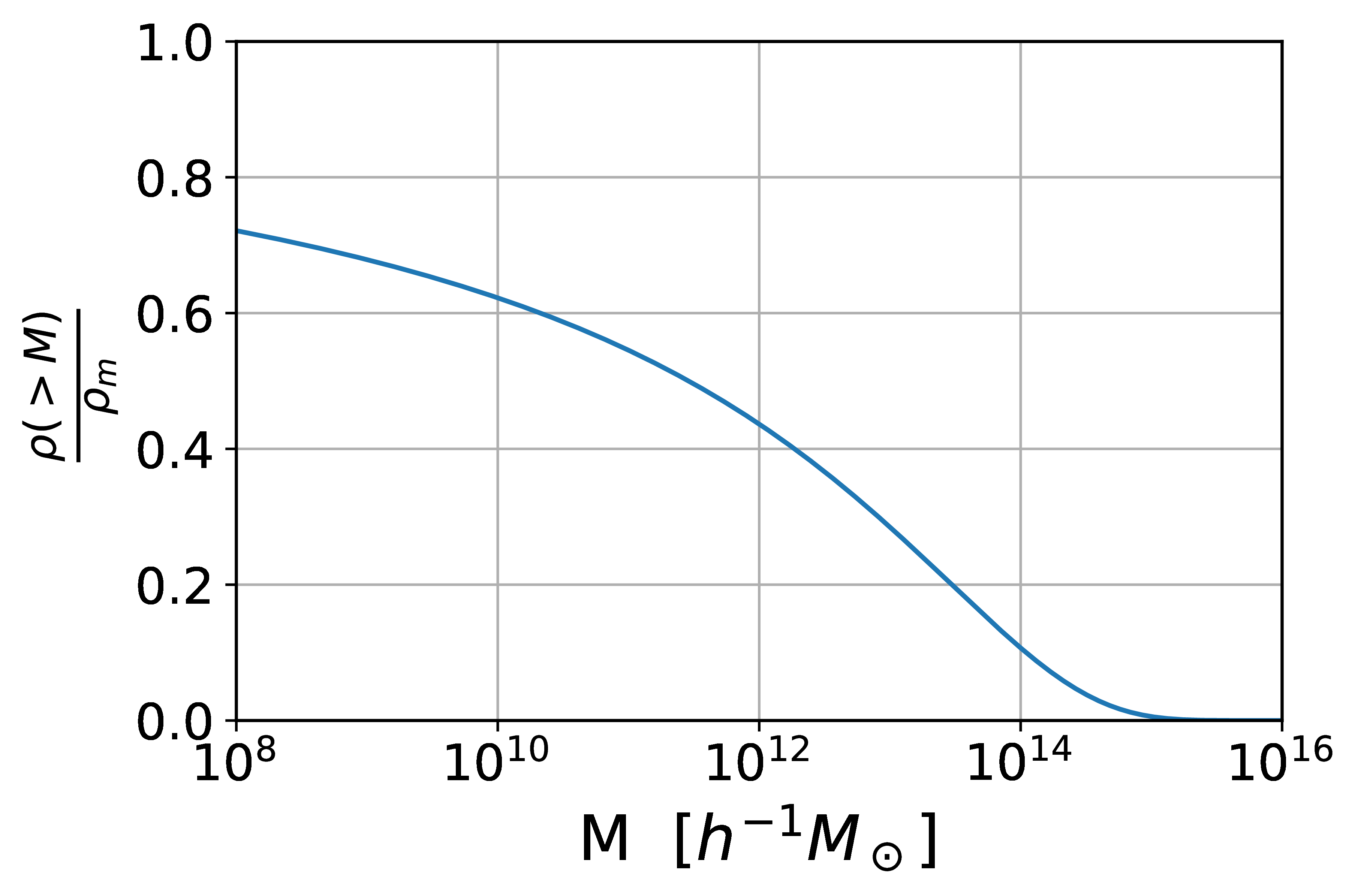}
    \caption{\textbf{Integrated mass in halos above mass M -}  Integration of the mass function of halos out to the splashback radius from  n-body calculations downward from the most massive divided by $\rho_m$.  The divide between luminous and non-luminous halos happens around $10^{10}\Msun$.  The IHM fraction would be 1 minus this fraction.
    \label{fig:halos}}
\end{figure}

A recent study by \cite{2020ApJ...903...87D} considers the mass within the second turnaround (splashback) radius of halos in simulations, a choice that matches the radius used in the definition of our halos. 
We chose to integrate their mass spectrum, as found in the Colossus toolkit \citep{2018ApJS..239...35D}, that includes particles within the radius that includes 90\% of the second turnaround points.
Integrating downward from the most massive halos (Figure~\ref{fig:halos}), one finds 59\% of the total mass is in halos down to $2 \times 10^{10} \Msun$ for $h = 0.7$.
Hence 41\% of mass lies in smaller halos or at large in sheets, filaments, and voids, a value most consistent with our $H_0 \sim$72 result, as shown in Figure~\ref{fig:Ox_H}. Establishing an accurate errorbar for this number is problematic at this time. 

\subsection{Local Group velocity}

The NAM models provide direct knowledge of the motion of the Local Group in the frame of the center of mass of the whole sample, including the external sources.  For the $H_0=73$ model, the present $SGX,SGY,SGZ$ motion is 
(\mbox{$-194$}, 292, \mbox{$-231$})~\kms\ or  
420~\kms\ into $SGL,SGB$ of (124\degr, \mbox{$-34$\degr}), galactic $l, b$ of (264\degr, 39\degr).   
Subtracting this vector from the CMB dipole leaves a residual velocity of 
240~\kms\ into $SGL,SGB$ (165\degr, \mbox{$-22$\degr}), or 
galactic $l, b$ of (293\degr, \mbox{$12$\degr}).  
It is not clear how much more motion comes from large scale structure beyond our external catalog, or unrepresented voids in the opposite direction \citep{2017NatAs...1E..36H} or from a more complete accounting of the masses in the region covered by the external catalog.

\subsection{Influence of external catalog}

If we remove the 44 external masses in our external catalog and hold all paths from the $H_0=73$ solution fixed except for the path of the Local Group, the direct effect of the external catalog on the Local Group can be seen.  The Local Group $SGX,SGY,SGZ$ motion is $V_{LG}$ =  (\mbox{$-161$}, 290, \mbox{$-234$})~\kms\ or  
406~\kms\ into $SGL,SGB$ of (119\degr, \mbox{$-35$\degr}), galactic $l, b$ of (259\degr, 40\degr).  Subtracting the Local Group motion with the external masses from the result of the previous section gives a Local Group motion arising from external masses into $SGX,SGY,SGZ$ of 
$V_{LG}^{ext}$ = (\mbox{$33$}, -2, -3) or 33~\kms\ into $SGL,SGB$ = (356\degr, -5\degr) or galactic $l,b$ = (141\degr, -4\degr), a direction into $SGX$ and the zone of avoidance.  Its low amplitude means that, apparently, the objects from around the sky between 8,000 and $16,000$ \kms\ mostly cancel one another and result in a very small net motion.

\subsection{Virgo Cluster} \label{sec:Virgo}

The nearby and well studied Virgo Cluster has a relatively poor $\chi = 2.15$ in the fiducial $H_0=73$ model (see Column 9 of Table~\ref{table:runs}).  The observed average redshift for the cluster is 1058~\kms\ in the Local Group frame (1150~\kms\ heliocentric) and its distance with the assumed zero point is 15.8~Mpc, resulting in a peculiar velocity of $-94$ \kms.
The fiducial model gives a peculiar velocity of $-111$~\kms, but to get so close in cz, Virgo was moved out to 16.3 Mpc, about 1$\sigma$ off.  When placed at its observed distance, the peculiar velocity is $-241$ \kms.
Models with a lower mass get better $\chi^2$, 
but the mass in the fiducial model already strains the lower mass limit.  
The model mass of the cluster is 
$3.2 \times 10^{14} (\frac{d}{15.8\mathrm{Mpc}}) \Msun$ while the virial theorem mass estimate is $6.0\pm 0.9 \times 10^{14} (\frac{d}{15.8\mathrm{Mpc}}) \Msun$ \citep{2020A&A...635A.135K}.  The $H_0=77$ cases reveal that higher $H_0$ can improves both cz and mass, but only slightly.

The peculiar velocity of Virgo is predominantly the component of the velocity of the Local Group in the Virgo direction because the radial component of the cluster's velocity is quite low in all cases.  Therefore, issues related to the path of cluster are probably not an important source of error.  One might suspect problems with the specific choice of path for the Local Group depending on tugs from nearer masses, but a look at our neighbors show all moving towards Virgo at nearly the same velocity.
A possibility is that the IHM density is biased \citep{1986ApJ...304...15B} to unusually low densities in this region rich in halos.  The overdensity to the Virgo Cluster of $\sim$2.0 in our fiducial model, could be as low as $1.5$ without an IHM and that would reduce the peculiar velocity by 25\%.

On the observational side, the $\chi$ could be improved with a lower observed velocity (unlikely) and/or a greater observed distance (more plausible).  The assumed statistical distance uncertainty is 3\% or 0.5~Mpc.  The depth of the cluster from fore to backside of the second turnaround surfaces (roughly the virial dimensions) is 4~Mpc, 8 times the distance uncertainty of the center.  The Virgo Cluster has structural complexity \citep{2007ApJ...655..144M, 2012ApJS..200....4F}.  Many galaxies are falling in today from the Southern Extension \citep{1984ApJ...281...31T, 2017ApJ...843...16K}.  M49, the brightest galaxy in the cluster, is near the geometric edge with its own entourage. 
The center of mass of the cluster, with an extensive and still unsettled halo, may not be precisely at the mean position of its galaxies.
If the simplistic description of Virgo as a single entity is relaxed, solutions could be found with reduced $\chi$ but great ambiguity.

\subsection{Centaurus Cluster}

\begin{figure}[ht]
    \centering
\includegraphics[width=0.9\linewidth]{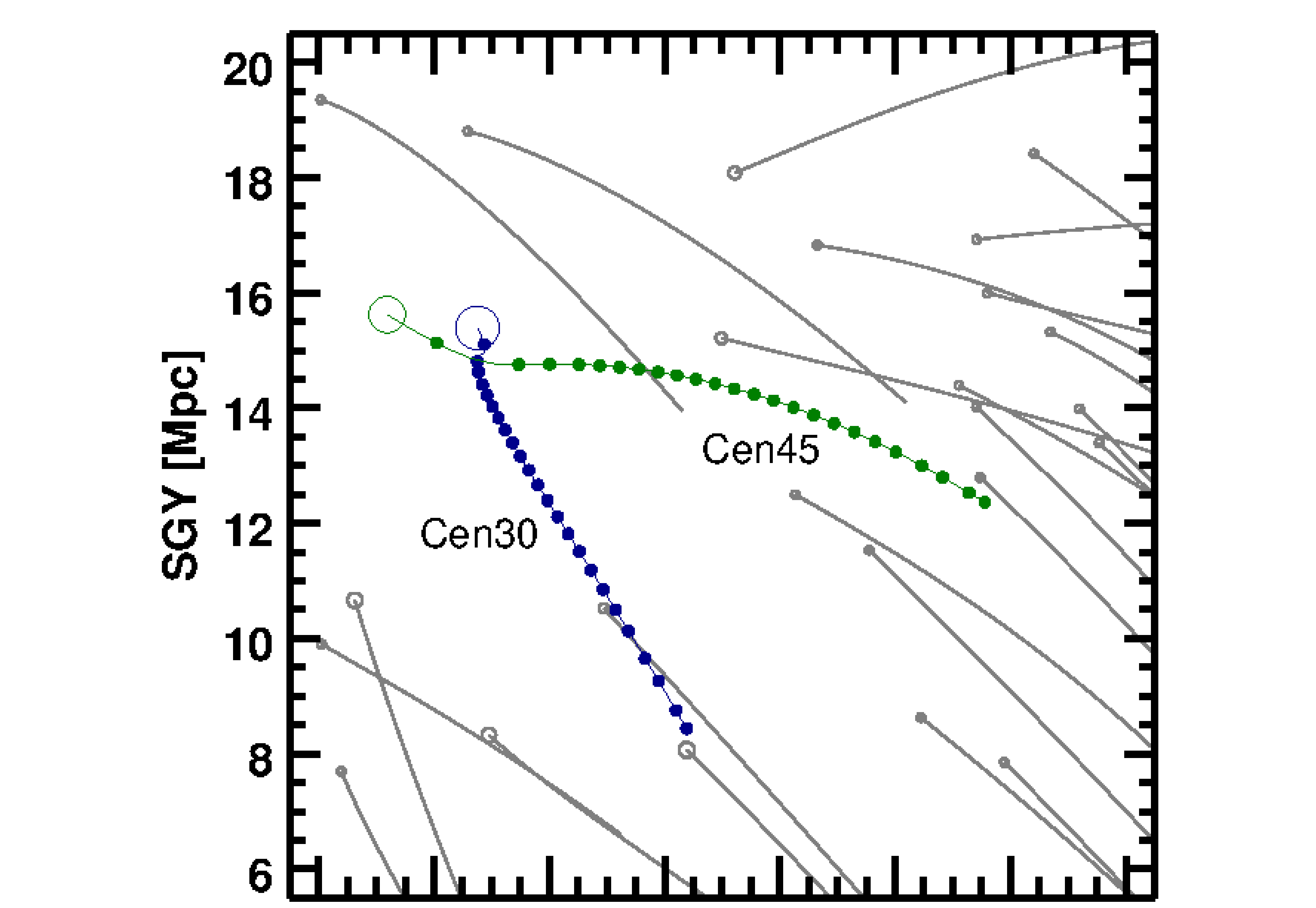}
    \includegraphics[width=0.9\linewidth]{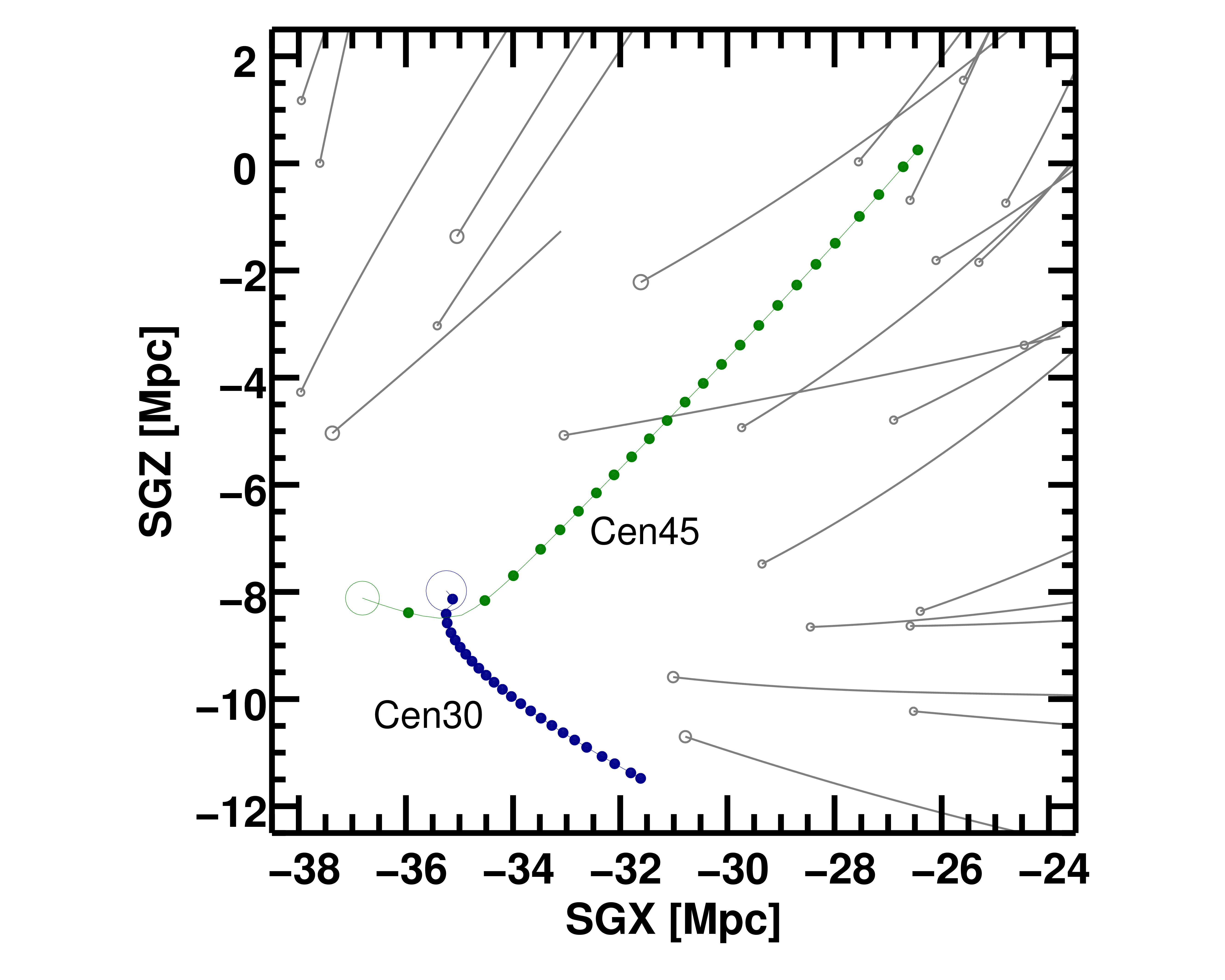}
    \caption{\textbf{Possible Paths for Cen30 and Cen40 Clusters} - The Centaurus Cluster is considered to be two clusters superimposed along the line of site.  Trajectories in comoving supergalactic coordinates have filled circles at 0.5 Gyr intervals and large open circles show the position at the present day.} 
    \label{fig:cen3045}
\end{figure}

Another cluster that defies a simple description is the Centaurus Cluster.  This cluster has been suspected to be two superposed clusters which have been dubbed Cen30 and Cen45 because one component peaks in velocity at $\sim$3,000~\kms\ and the other peaks at $\sim$4,500~\kms\ \citep{1986MNRAS.221..453L, 1997A&A...327..952S}.  
At first, we retained it as a single cluster in our runs at $d_o = 37 \pm 1.1$ Mpc and with $cz$=3,142 \kms.  
The model output distances were all around 43 Mpc, which is unacceptable.  
We therefore broke it up, guided by its double-peaked velocity distribution, into the two suspected components.
Based on 43 mainly fundamental plane and spiral luminosity-linewidth distances,
Cen30 was placed at 37.8~Mpc with $cz$ = 2,825 \kms\ while with 23 distances Cen45 was placed at 45.5 Mpc with $cz$=4,306 \kms, where Cen30 is about twice as massive as Cen45.  The separation between the two components based on all Cosmicflows-3 measures is $8\pm2$~Mpc.  It is to be noted, though, that surface brightness fluctuation measures found the two components to be at similar distances within the uncertainties \cite{2003A&A...410..445M, 2005A&A...438..103M}.
A special search of solutions was made by repeatedly `backtracking' their paths from the present positions and redshifts to find a possible strong interaction, but the separation of about 8 Mpc is too large.
Interacting scenarios with separations of about 2 to 3 Mpc, at best, were found that provide enough positive peculiar velocity to Cen30.

Figure~\ref{fig:cen3045} shows such an example of a close interaction in which Cen45 flies by Cen30 and then goes out to $\sim$2 Mpc beyond it.  Cases with greater separation were found, but they required Cen45 having an earlier interaction with yet another halo.  Although possible, we have refrained from using such complicated scenarios. 
Due to the large uncertainty of such models related to the ambiguity in the relative distances of Cen30 and Cen45, we have removed both of these halos from all statistics in this study.

\section{Flow Patterns within 100 Mpc}
\label{sec:flows}

\begin{figure}[ht]
    \centering
    \includegraphics[width=0.47\textwidth]{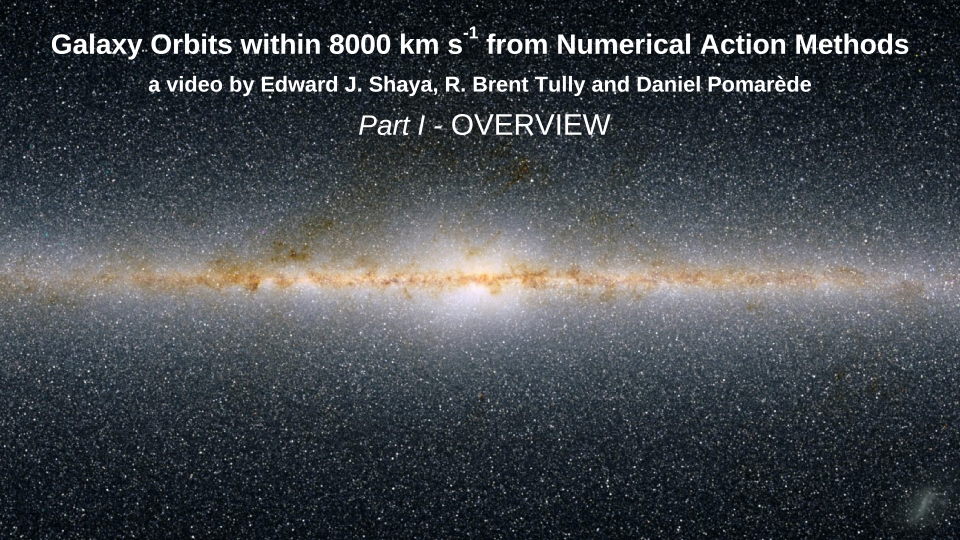}
    \caption{\textbf{Overview video -} The high-resolution video in UHD-4K format is available online for viewing and download at \href{https://vimeo.com/pomarede/nam8k-overview}{https://vimeo.com/pomarede/nam8k-overview}}
    \label{fig:video_overview}
\end{figure}

\begin{figure}[ht]
    \centering
    \setlength{\fboxsep}{0pt}\fbox{\includegraphics[width=0.47\textwidth]{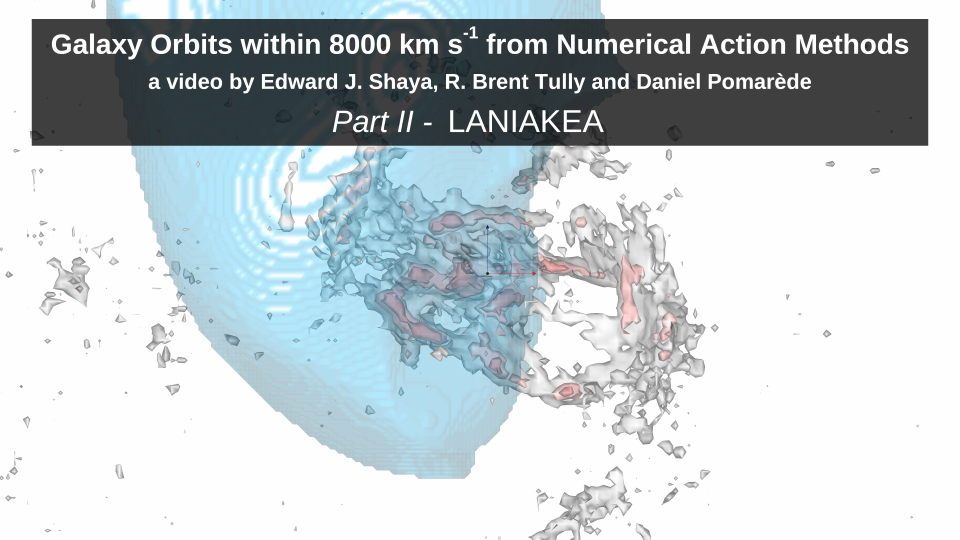}}
    \caption{\textbf{Video visualization of the orbits in Laniakea.} The high-resolution video in UHD-4K format is available online at \href{https://vimeo.com/pomarede/nam8k-laniakea}{https://vimeo.com/pomarede/nam8k-laniakea}}
    \label{fig:video_laniakea}
\end{figure}

\begin{figure}[ht]
    \centering
    \setlength{\fboxsep}{0pt}\fbox{\includegraphics[width=0.47\textwidth]{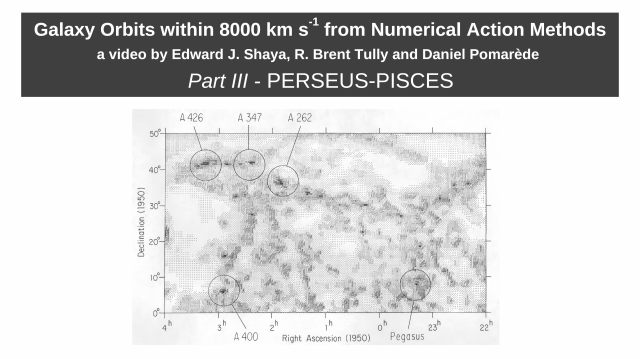}}
    \caption{\textbf{Video visualization of the orbits in Perseus-Pisces.} The high-resolution video in UHD-4K format is available online at \href{https://vimeo.com/pomarede/nam8k-perseus-pisces}{https://vimeo.com/pomarede/nam8k-perseus-pisces}}
    \label{fig:video_perseuspisces}
\end{figure}

\begin{figure}[ht]
    \centering
    \setlength{\fboxsep}{0pt}\fbox{\includegraphics[width=0.47\textwidth]{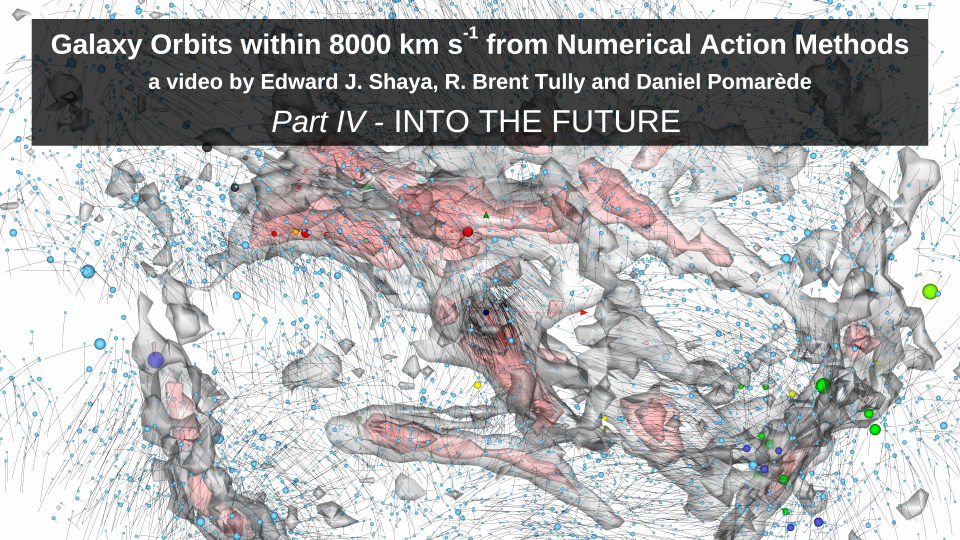}}
    \caption{\textbf{Video of the Future.} The high-resolution video in UHD-4K format is available online at \href{https://vimeo.com/pomarede/nam8k-intothefuture}{https://vimeo.com/pomarede/nam8k-intothefuture}}
    \label{fig:video_intothefuture}
\end{figure}

\begin{figure*}[ht]
    \centering
    \includegraphics[width=.9\textwidth]{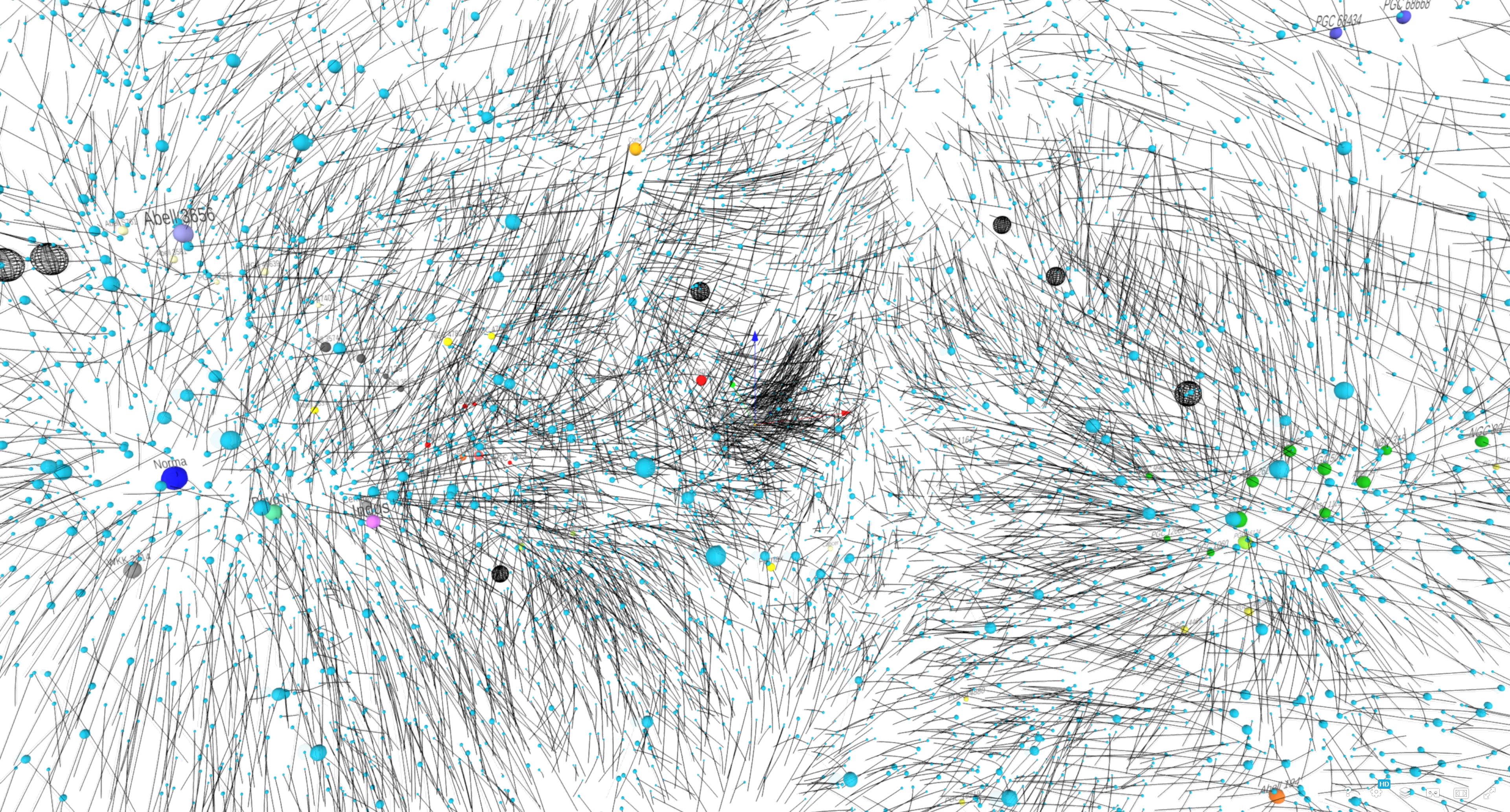}
    \caption{\textbf{Overview visualization displaying the entire sample of 9,719
    model objects and 44 externals.} Start interaction at
    \href{https://bit.ly/NAM8Koverview}{https://bit.ly/NAM8Koverview}. The red, green, blue 1,000 km/s-long arrows emanating from the origin indicate the three cardinal axes of the Supergalactic coordinate system. }
    \label{fig:Sketchfab_overview}
\end{figure*}

\begin{figure*}[ht]
    \centering
    \includegraphics[width=0.9\textwidth]{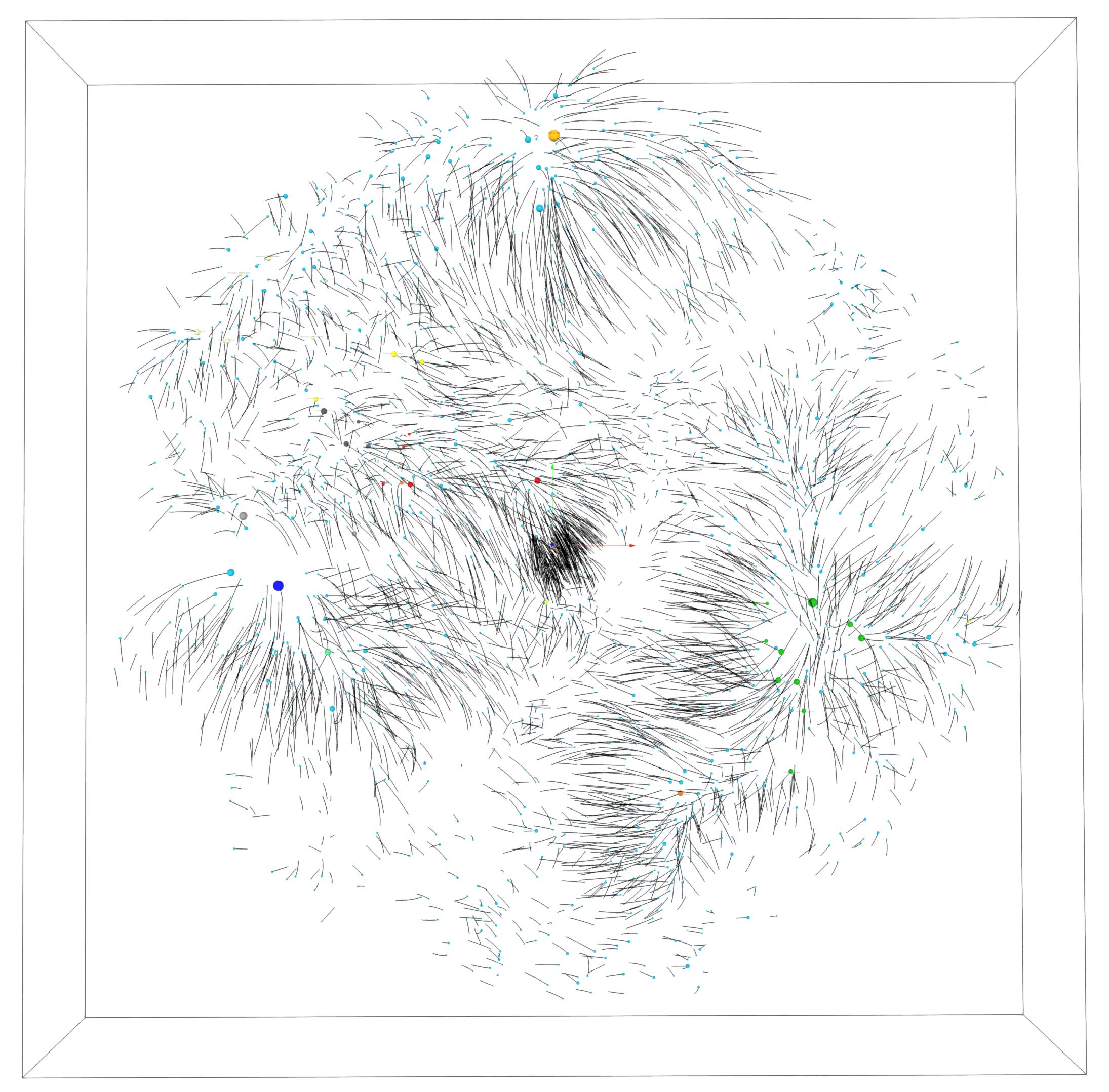}
    \caption{\textbf{Visualization of the Supergalactic Equator plane}, where the
    final positions of objects is restricted to the slice $-20<SGZ<+20$ Mpc. Start
    interaction at \href{https://bit.ly/NAM8Kequator}{https://bit.ly/NAM8KFigequator}.}
    \label{fig:Sketchfab_supergalacticequator}
\end{figure*}

Our discussion takes frequent recourse to the accompanying group of four videos\footnote{The four videos are narrated by Prof. R. Brent Tully.} and four interactive models; see animated Figures~\ref{fig:video_overview},~\ref{fig:video_laniakea},~\ref{fig:video_perseuspisces},~\ref{fig:video_intothefuture} and interactive Figures~\ref{fig:Sketchfab_overview},~\ref{fig:Sketchfab_supergalacticequator},~\ref{fig:Sketchfab_greatattractor},~\ref{fig:Sketchfab_perseuspisces}\footnote{The videos and interactive visualizations can also be accessed at the following address: \href{http://irfu.cea.fr/nam8k}{http://irfu.cea.fr/nam8k}}.  
In the opening overview video (Figure~\ref{fig:video_overview}) and interactive model (Figure~\ref{fig:Sketchfab_overview}), the first thought of the view of 10,000 orbits is likely to be of a bad hair day. A noticeable gap runs horizontally.  This poorly represented part of the plot is associated with the zone of obscuration of the Milky Way and will be remarked frequently depending on the orientation of views.  Elements of the model do lie in this zone sufficient to maintain roughly constant density, as described in \S 2, but they are numerically few.  The eye also sees a great concentration of orbits near the center.  Near to home, within 10~Mpc, there are an abundance of (mostly light weight) galaxies with accurately known distances from the tip of the red giant branch method \citep{2021arXiv210402649A}.

At intervals throughout the videos, one sees the superposition of density contours on the orbits.  These surfaces are derived from a quasi-linear Wiener filter analysis of {\it Cosmicflows-3}.  The methodology is described in an earlier analysis of {\it Cosmicflows-2} data \citep{2018NatAs...2..680H}.  These contours provide a reference frame to aid in orientation, appreciation of scale, and to show target destinations for flow patterns.  The quasi-linear maps are chosen over other options because of the excellent resolution that they afford.  The quasi-linear analysis with {\it Cosmicflows-3} data will be described in detail by Hoffman et al. (in preparation).

In the big picture, it is immediately evident that there are three overarching flow pattern zones within 8,000~\kms, which will be referred to as Laniakea, Perseus-Pisces, and Great Wall, as seen in the interactive Figure~\ref{fig:Sketchfab_supergalacticequator}. Each will be discussed in turn.  The first two are reasonably contained within the boundary of the study.  Only the closest parts of the Great Wall are being seen so the current description of flows in the region of that structure is perfunctory. Another dominant overview feature is the relative absence of mass elements in the sector at negative $SGY$ (the sky south of the Galactic plane) and negative $SGX$.  See in particular the interactive model restricted to $\pm20$~Mpc about $SGX=0$ (Interactive Figure~\ref{fig:Sketchfab_supergalacticequator}).

\subsection{Laniakea}
\label{sec:laniakea}

The concept of the Laniakea Supercluster was introduced by \citet{2014Natur.513...71T}.  The Milky Way lies within its embrace.  The major over density constituents, strongly linked together, are the classical Local Supercluster \citep{1953AJ.....58...30D}, the so-called Great Attractor \citep{1987ApJ...313L..37D}, and the Pavo$-$Indus filament \citep{1996MNRAS.283..367D}.  In all, there are 18 Abell clusters within the envelope of Laniakea if one counts the supplemental entries in the list of \citet{1989ApJS...70....1A} and the Virgo Cluster.  As must be expected, the overdense structures are encased in voids that take up most of the volume of Laniakea.

The pattern of orbits in the region of the Local Supercluster were discussed extensively by \citet{2017ApJ...850..207S} and the main points are summarized here. The dominant patterns within this nearby region are flows out of the Local Void and toward the Virgo Cluster. The entire region is participating in a flow toward the Great Attractor.  Specifically, (1) essentially all galaxies north of the supergalactic equator (the sector of the Local Void), those with positive values of $SGZ$ in supergalactic coordinates, are moving toward negative $SGZ$, whereas galaxies south of the supergalactic equator have disorganized $SGZ$ motions. Also, (2) essentially all galaxies across the region have motions toward negative $SGX$ and positive $SGY$, the direction toward the Great Attractor and the apex of the CMB dipole.

\begin{figure*}[ht]
    \centering

    \includegraphics[width=0.9\textwidth]{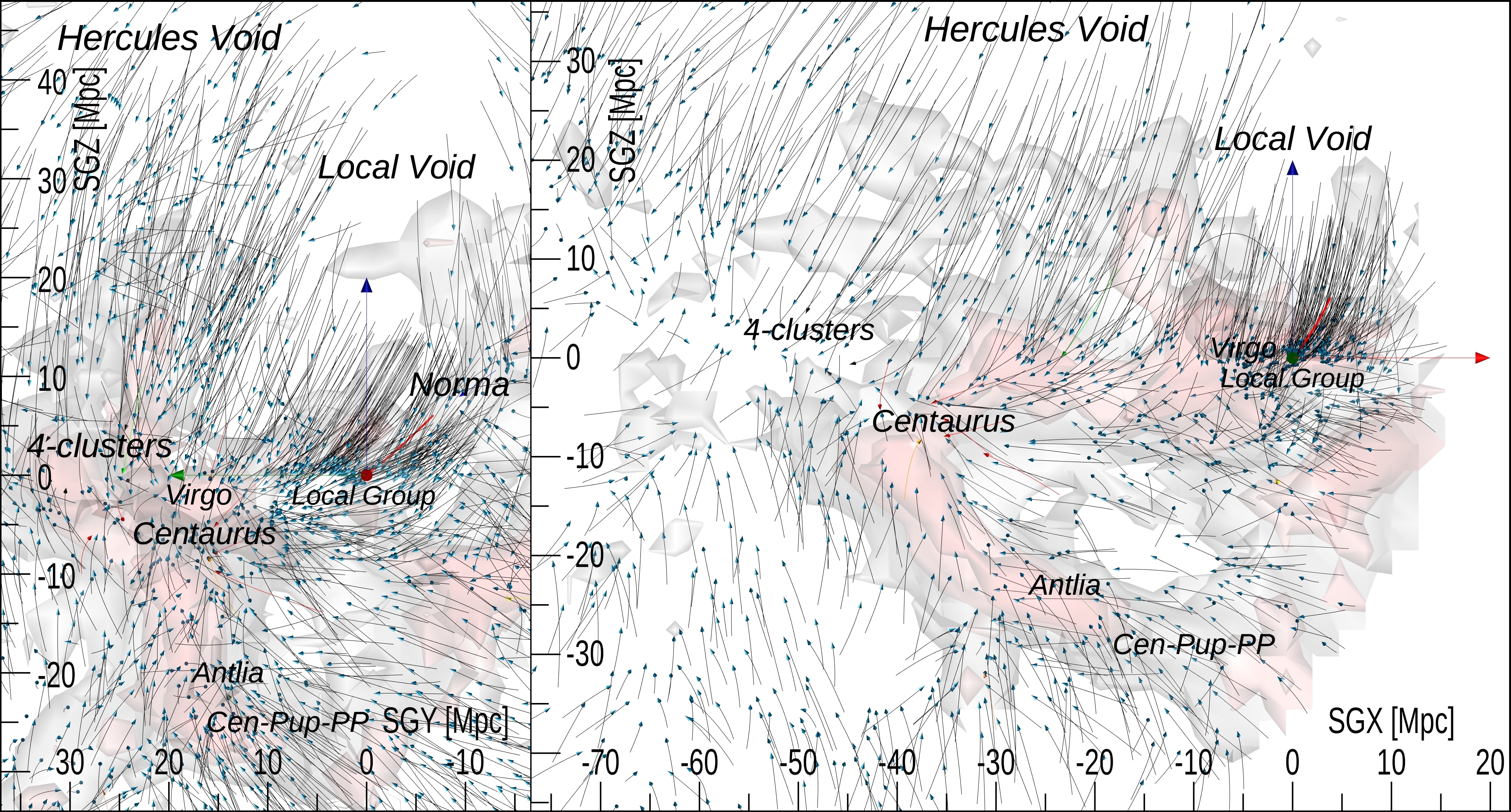}
    \caption{\textbf{Orthogonal views of orbits in the Great Attractor region.}  At left is a view from $-SGX$ and at right is a view from $-$SGY. The orbit of the Local Group is in red, terminating at the origin marked by orientation arrows: red toward $+SGX$, green toward +SGY and blue toward +SGZ.  Throughout this presentation, the orientation arrows are 20~Mpc in length. Isodensity contours of the quasi-linear model are shown faintly.}
    \label{fig:ga_xy}
\end{figure*}

\begin{figure*}[ht]
    \centering
    \includegraphics[width=0.9\textwidth]{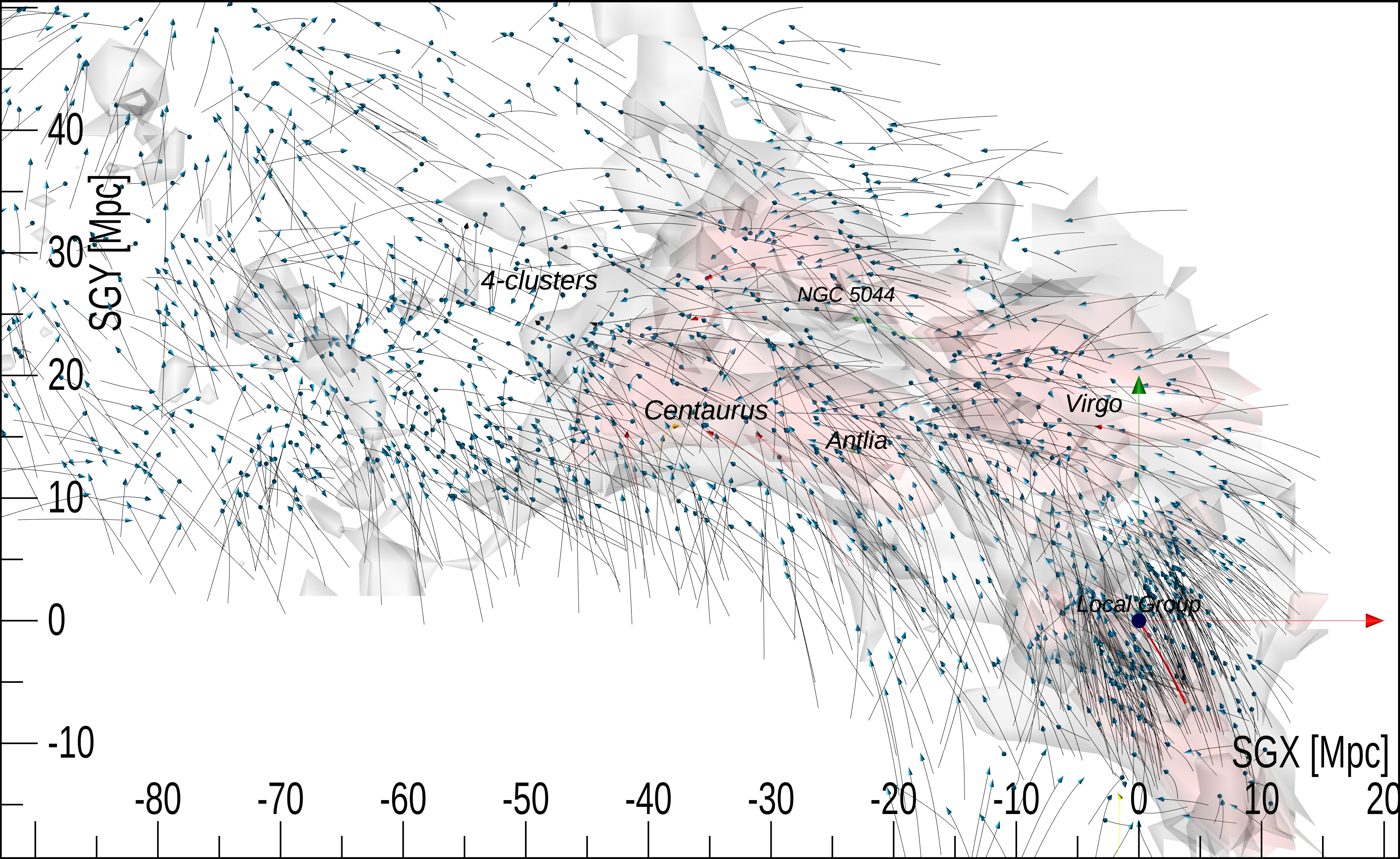}
    \caption{\textbf{Orbits in the Great Attractor region -} viewed from the north supergalactic pole (+SGZ). }
    \label{fig:ga_z}
\end{figure*}

We retain the nomenclature ``Great Attractor" but what does this phrase mean?  We take it to refer to the rough terminus of flow patterns within the Laniakea envelope.  It is not a well defined location.

An overview of the Great Attractor region is shown in the two panels of Figure~\ref{fig:ga_xy}.  The scenes are approximate extracts from the Part-1 (Overview, see Figure~\ref{fig:video_overview}) video at 5:50 and the Part-2 (Laniakea, see Figure~\ref{fig:video_laniakea}) video at 3:10.  The overwhelmingly dominant pattern is the downward flow of orbits toward negative $SGZ$ for galaxies above the supergalactic equator at $SGZ=0$.  This pattern is the same as was seen in the Local Supercluster study, now exhibited on a much larger scale.  The space at positive $SGZ$ is the domain of the Local Void blending into the Hercules Void  \citep{2019ApJ...880...24T}.  There is also a trend of motions upward from galaxies at negative $SGZ$ toward $SGZ=0$ but with less coherence across the full domain of these plots.  Closer attention is required to isolate patterns.

\begin{figure*}[ht]
    \centering
    \includegraphics[width=.9\textwidth]{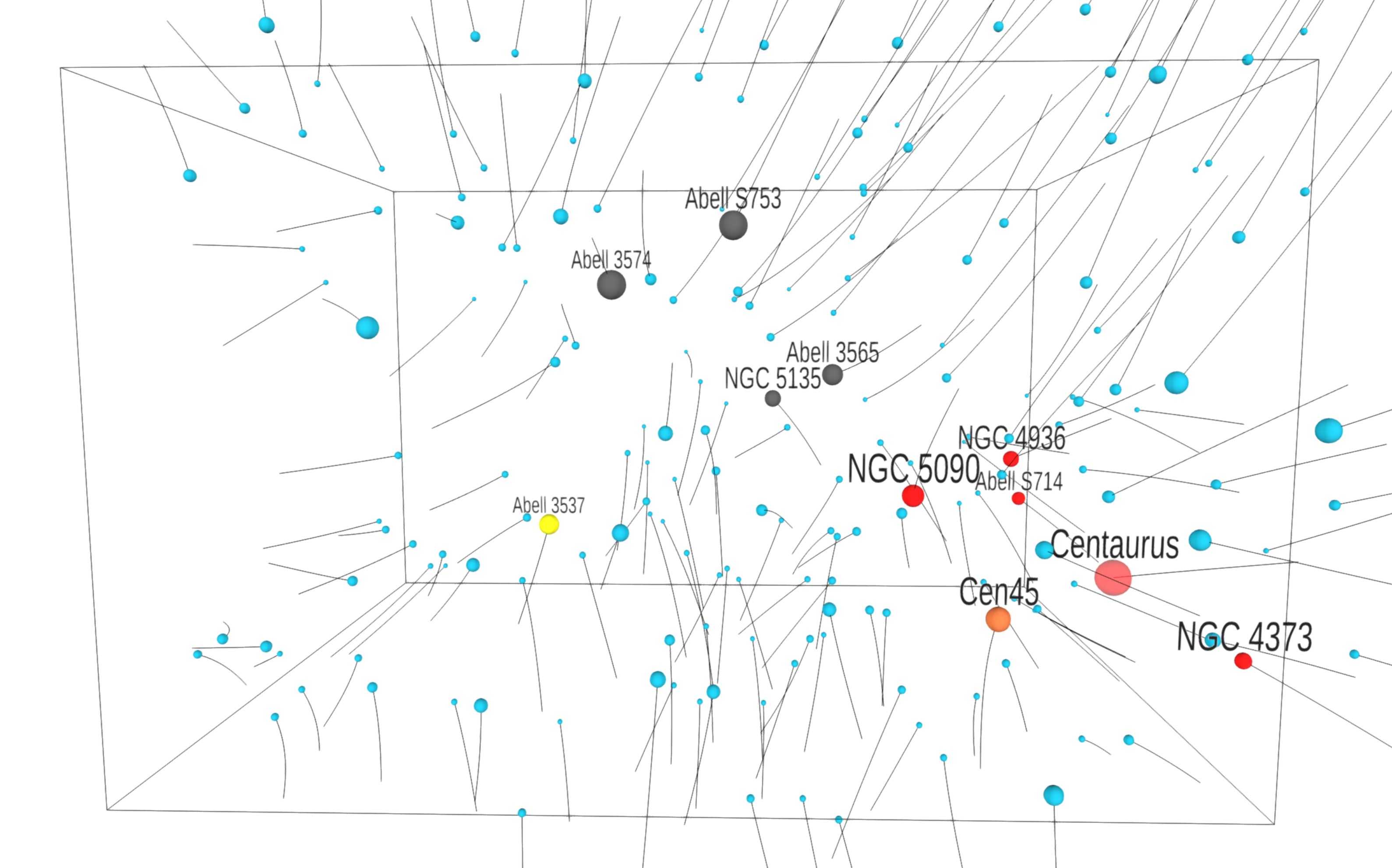}
    \caption{\textbf{Zoom visualization of the Great Attractor region.} Start 4D
    interaction at \href{https://bit.ly/NAM8KGA}{https://bit.ly/NAM8KGA}. In this visualization of the n-body solution, the time evolution is run on the basis of one second for one Gyr.}
    \label{fig:Sketchfab_greatattractor}
\end{figure*}

A prominent feature of local peculiar motions is a flow toward $-SGX$ in the Local Group vicinity, and it is seen in the right panel of Figure~\ref{fig:ga_xy} and in Figure~\ref{fig:ga_z} (see the Part-2 video at 2:45, Figure~\ref{fig:video_laniakea}) that this pattern persists from $SGX = +10$ Mpc to $SGX = -45$ Mpc
before breaking down.  Detailed attention suggests a convergence of flows roughly centered between the three clusters A3565, A3574, and S753 at roughly $SGX = -55$~Mpc, as seen in the interactive Figure~\ref{fig:Sketchfab_greatattractor}.  These entities are part of what was called the 4-cluster strand by \citet{2013AJ....146...69C}. A3565 and A3574 are each $\sim$8~Mpc from the more central S753, separations comparable to each of the individual turnaround radii about these clusters.  It is plausible that these three clusters are destined to merge although this scenario is not favored in our discussion in \S\ref{sec:future}.  While there is a convergence of flow lines in their proximity, it is a bit of a mystery that this location is not particularly prominent in the quasi-linear density maps. 

Beyond the 4-clusters area flows are confused.  The convergence of flows toward the supergalactic equator is retained: toward negative $SGZ$ from above the plane and toward positive $SGZ$ from below the plane.  However there is disorganization in the lateral $SGX$ motions.  We are approaching the 8,000~\kms\ edge of the orbit reconstruction volume here.  External tides are only crudely represented, and not far beyond in this sector is the Shapley Concentration, the densest collection of rich clusters within $0.1c$ \citep{1992ApJ...388....9T}.
Evidently, the fringes of the influence of Laniakea Supercluster are being probed.  Deeper surveys are required for a reliable picture.

\begin{figure*}[ht]
    \centering
    \includegraphics[width=0.9\textwidth]{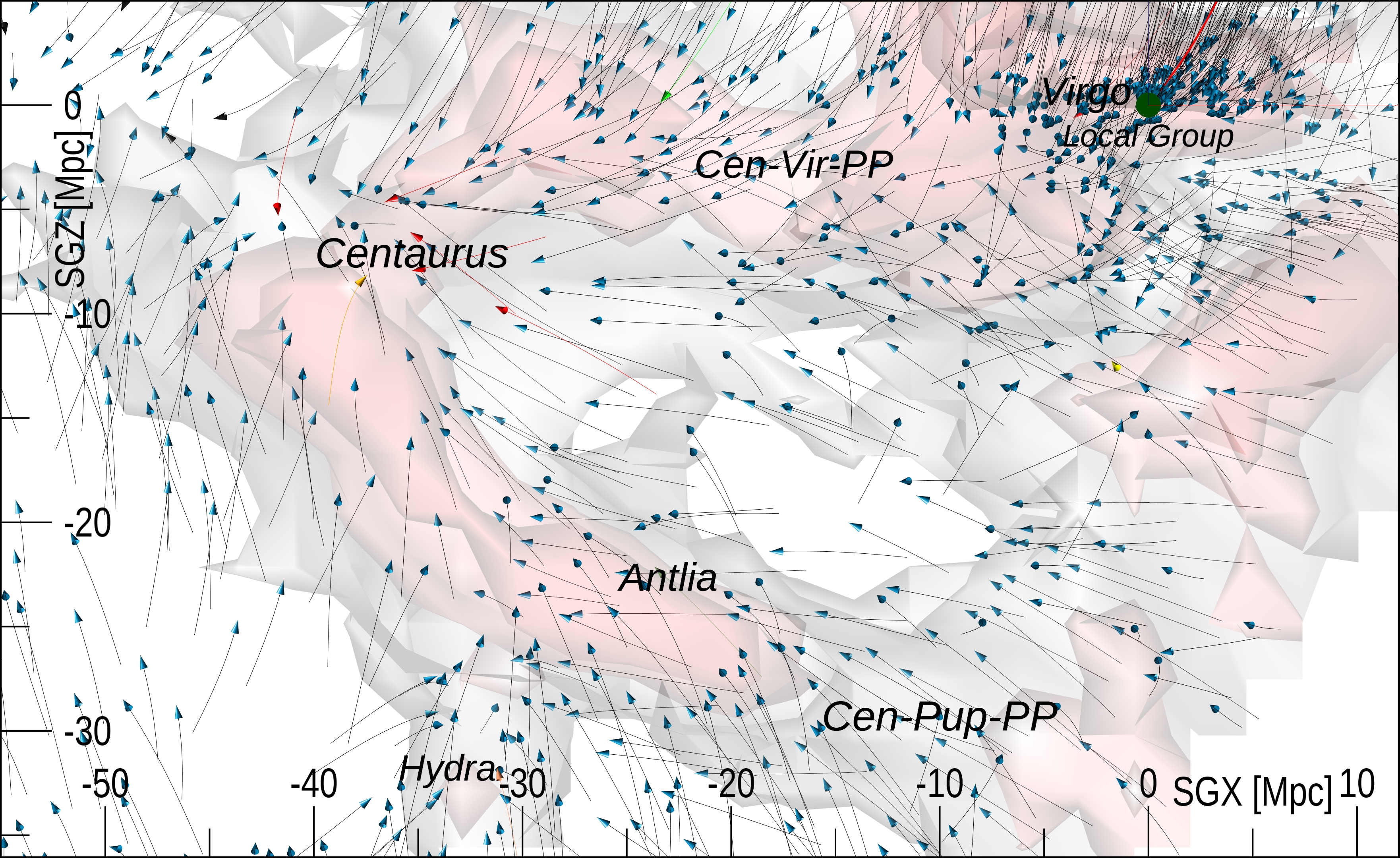}
    \caption{\textbf{Coherent flow pattern through the region of the Antlia Cluster to the region of the Centaurus Cluster.} This
    Antlia strand is a part of the more extensive Centaurus-Puppis-PP filament.}
    \label{fig:antlia}
\end{figure*}

Returning to the core region of the Great Attractor, the most prominent feature {\it off} the supergalactic equator is a structure running from the Centaurus Cluster to the Antlia Cluster at negative $SGZ$, slanting to positive $SGX$ but at almost constant $SGY$ (see Figure~\ref{fig:antlia}).  This structure was called the Hydra Wall by \citet{1998lssu.conf.....F} but the Hydra Cluster is to the background.  It is called the Antlia strand by \citet{2013AJ....146...69C}; one of five filamentary structures that converge on the Centaurus Cluster (two of the others have already been given attention; the 4-clusters strand and the planar Local Supercluster that runs between the Virgo and Centaurus clusters).  The flow pattern running up from Antlia Cluster through Centaurus Cluster is coherent and pronounced.  As seen in Figure~\ref{fig:antlia} (the Part-3 Perseus-Pisces video at 5:00, Figure~\ref{fig:video_perseuspisces}),
if followed beyond Antlia toward positive $SGX$ there is a structure that extends all the way over to the Perseus-Pisces complex.  This feature will be given attention at a further point in the discussion.

\begin{figure}[ht]
    \centering
    \includegraphics[width=0.47\textwidth]{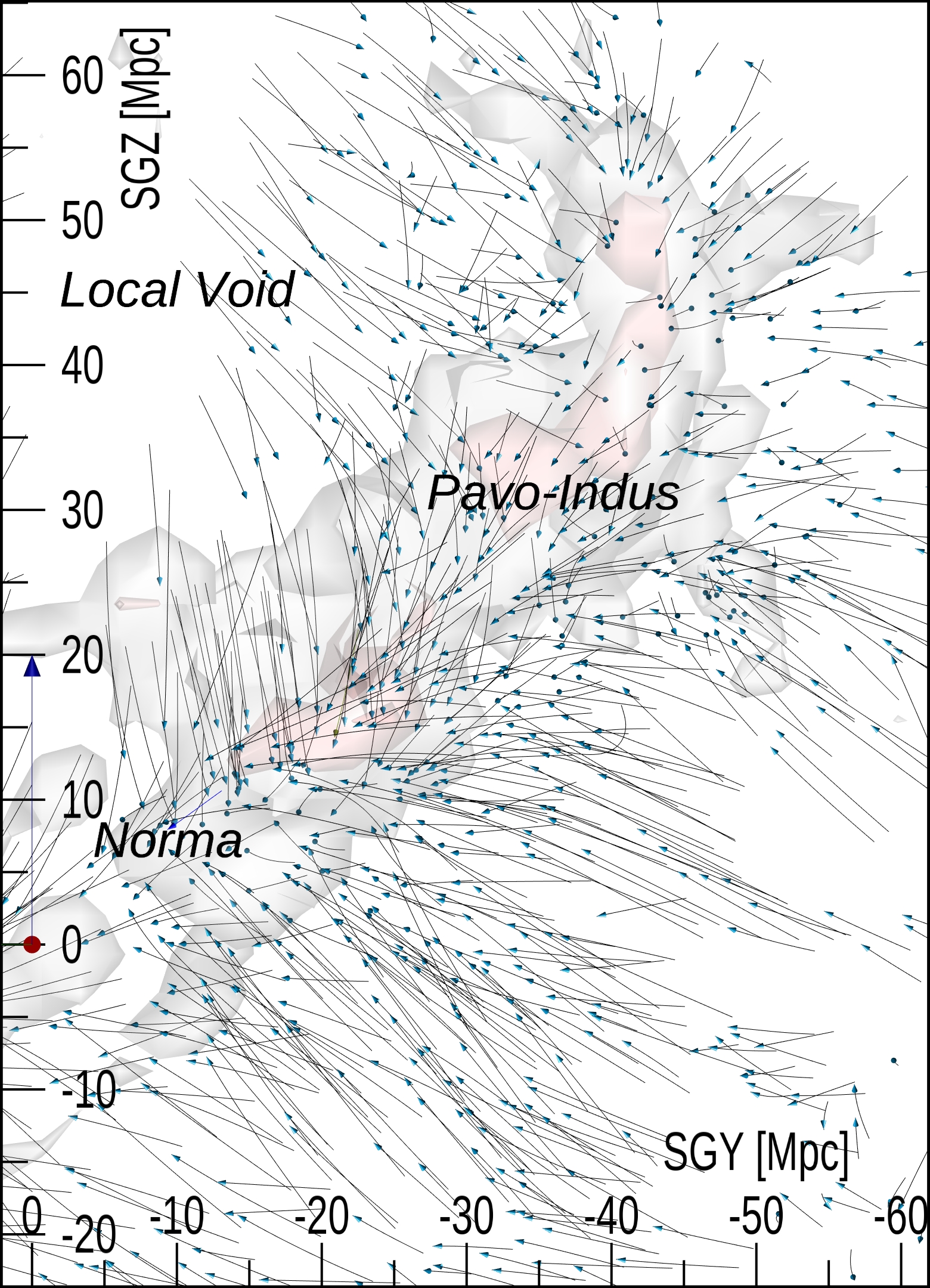}
    \caption{\textbf{The Pavo-Indus filament -} a prominent structure that rises above the supergalactic equator to embrace the Local Void.}
    \label{fig:pavoindus}
\end{figure}

The really major appendage off the Great Attractor core of Laniakea is the Pavo-Indus filament running from the Norma Cluster to high supergalactic latitudes. \cite{1998lssu.conf.....F} called this structure the Centaurus Wall.  Some ambiguity arises regarding its connection with the Great Attractor region because of the intervening zone of obscuration \citep{2000A&ARv..10..211K, 2006MNRAS.369.1131R}.  The suggestion that Pavo-Indus is connected to the Perseus-Pisces complex by \citet{1996MNRAS.283..367D} has more recently been confirmed by \citet{2017ApJ...845...55P} through a feature called the Arch.  Pavo-Indus is a cupped sheet-like structure, a curved partial wall enclosing the Local Void \citep{2019ApJ...880...24T}.  Flow patterns illustrated in Figure~\ref{fig:pavoindus} (Part-2 Laniakea video at 5:00, Figure~\ref{fig:video_laniakea}) are very coherent, rushing toward the surface of the sheet from both sides and downward toward Norma Cluster and the Great Attractor core.

\subsection{Perseus-Pisces}
\label{sec:pp}
\begin{figure}[ht]
    \centering
    \includegraphics[width=0.47\textwidth]{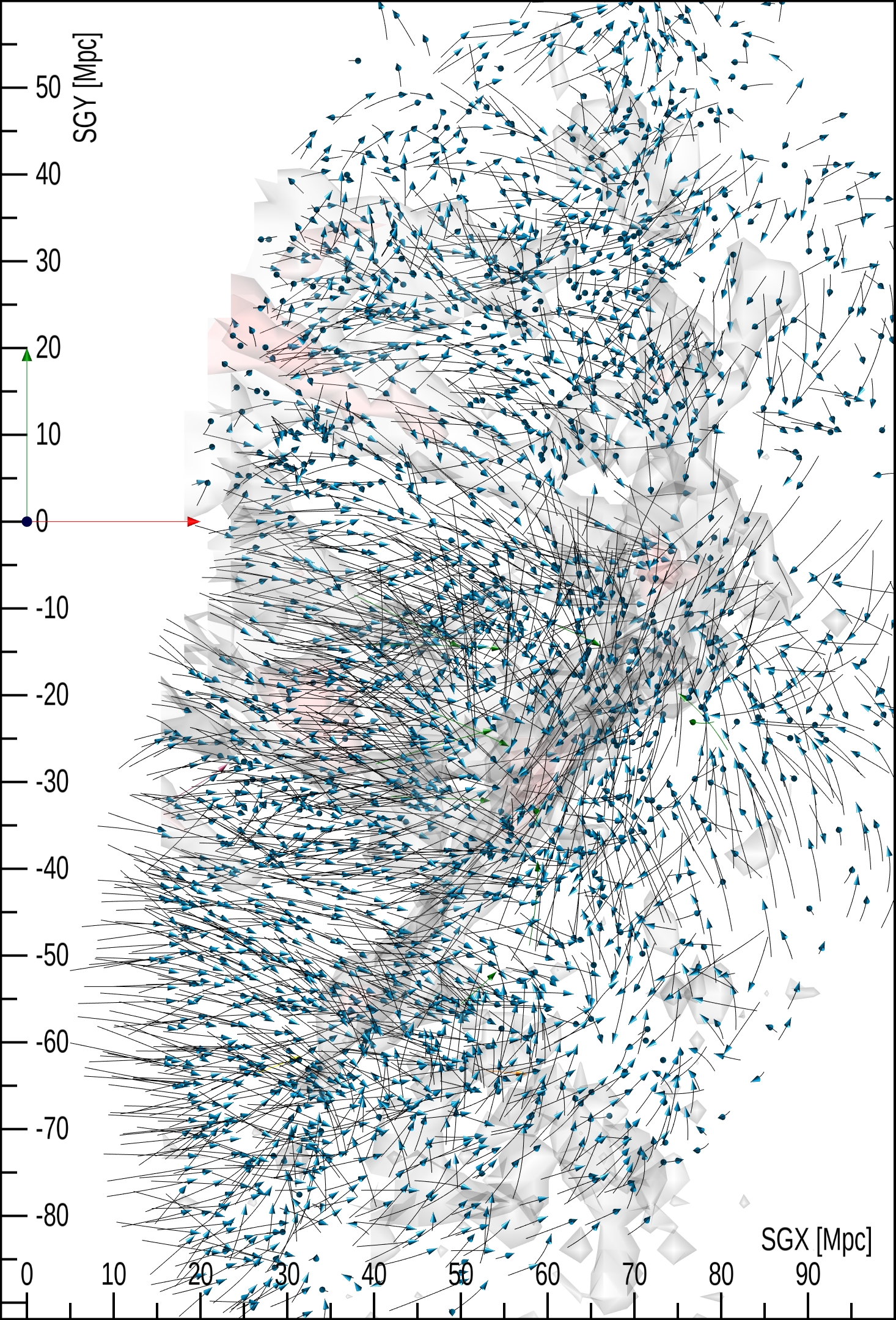}
    \caption{\textbf{Flows toward the spine of the Perseus-Pisces filament -} viewed from the supergalactic north pole (+SGZ).  The Local Group lies at the origin marked by the red-green-blue orientation arrows.}
    \label{fig:pp+sgz}
\end{figure}

\begin{figure*}[ht]
    \centering
    \includegraphics[width=1.0\textwidth]{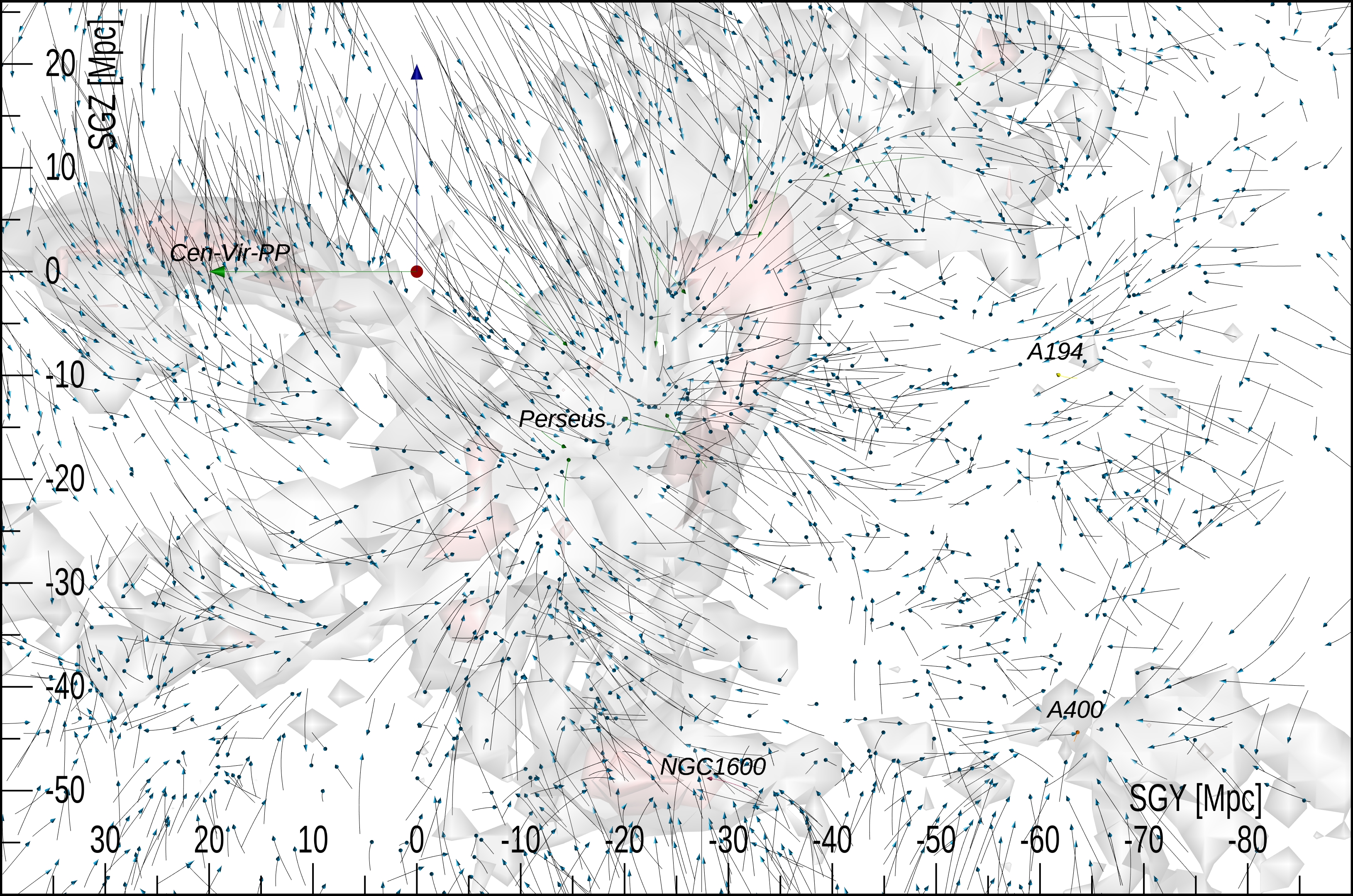}
    \caption{\textbf{An orthogonal view of the Perseus-Pisces region looking in from $\mathbf{-SGX}$.}  The Perseus Cluster is a focus of flow patterns.  The A400 Cluster is a secondary focus of flows.  The NGC~1600 and Cen-Vir-PP features are parts of bridges to Laniakea Supercluster.}
    \label{fig:pp-sgX}
\end{figure*}

Our vantage point within Laniakea Supercluster is very near the boundary with the Perseus-Pisces complex.  The full extent of this entity remains to be defined because of the limitations in distance of current data.\footnote{We resist calling the Perseus-Pisces complex a supercluster because we may not be capturing the full extent of this structure.  Following from the discussion of Laniakea \citep{2014Natur.513...71T} we reserve the nomenclature of ``supercluster" to refer to the full gravitational basin of flow lines jointly decoupled from the cosmic expansion.  The full extent of the Perseus-Pisces structure remains to be defined.}  However the flow toward the spine of the Perseus-Pisces filament is dramatic, as seen at 6:32 of the Part-1 Overview video (Figure~\ref{fig:video_overview}, see also Figure~\ref{fig:pp+sgz}).  There is a coherent fore-side flow away from us toward positive $SGX$ out of the Local Void. There is evident backside flow in our direction toward negative $SGX$ although with less fidelity given the proximity to the sample boundary.  The Perseus Cluster is a pronounced focus of flows as can be seen at either 7:00 in the Part-1 Overview video (Figure~\ref{fig:video_overview}) or at 1:03 in the Part-3 Perseus-Pisces video  (Figure~\ref{fig:video_perseuspisces}, see also Figure~\ref{fig:pp-sgX}).

The Abell~400 Cluster is seen as a secondary focus of flows.  Not too much should be said about this feature because it is in proximity to the sample boundary.  \citet{2017ApJ...845...55P} have referred to structure in this region as the Funnel.  A case can be made that it is part of the extremely extensive South Pole Wall \citep{2020ApJ...897..133P}.

The features labeled NGC~1600 and Cen-Vir-PP are parts of filaments running toward us and connecting to our Laniakea Supercluster.  These filamentary links will be given attention later in the discussion.

\subsection{Great Wall}

The Great Wall \citep{1986ApJ...302L...1D} is roughly confined to an $SGX-SGZ$ plane at constant $SGY$ that intersects our 8,000~\kms\ volume at its nearest part in the vicinity of the Coma and A1367 clusters.  Very important parts, such as its extension to the major Hercules Cluster, are external to our study region.  There are coherent flows from the foreground toward positive $SGY$ and the Great Wall (8:16 in the Part-1 Overview video Figure~\ref{fig:video_overview}) arising out of the Hercules Void, the separator between Laniakea and the Great Wall \citep{2019ApJ...880...24T}.  Whatever motions arise from the background are lost to this study. 

\subsection{Supercluster Connections}

The most impressive bridge between the Laniakea and Perseus-Pisces gravitational basins was called the Centaurus-Puppis-PP filament by \citet{2017ApJ...845...55P}.  See at 5:00 in the Part-3 Perseus-Pisces video (Figure~\ref{fig:video_perseuspisces}). At the Laniakea end, near the Great Attractor core, this filament manifests as the Antlia strand that was discussed in \S\ref{sec:laniakea}.  At the Perseus-Pisces end, the filament is picked up at the NGC~1600 knot mentioned in \S\ref{sec:pp} and seen in Figure~\ref{fig:pp-sgX}. In between, this filament passes close enough for the very coherent flow pattern to have been mapped in detail in the numerical action orbit study of \citet{2017ApJ...850..207S}. As has become familiar with such bridges between potential wells, there is a point along the filament where there is a flow reversal \citep{2017ApJ...845...55P}.  In this case, NGC~1600 is within the embrace of Perseus-Pisces, with the reversal occurring somewhat toward negative $SGX$ of the group containing this galaxy.  

A second very evident bridge filament was called Centaurus-Virgo-PP by \citet{2017ApJ...845...55P}  because the band goes from Centaurus Cluster through the Virgo Cluster then onward to the Perseus Cluster.  Like the previously discussed Cen-Pup-PP filament, it is not well known because parts are hidden by the zone of obscuration.  As the filament enters within the dominant influence of the Perseus-Pisces gravitational domain the flow pattern orients toward the Perseus Cluster as seen in Figure~\ref{fig:pp-sgX} and the Part-3 Perseus-Pisces video segment at 1:03 (see animated Figure~\ref{fig:video_perseuspisces}). 

The Part-3 Perseus-Pisces video at 6:45 also gives attention to a less well defined structure (in parts, because of the considerable distance and its concealment by obscuration) that \citet{2017ApJ...845...55P} called the Centaurus-Arch-PP filament.  This feature is the high $SGZ$ cap on the Local Void \citep{2014Natur.513...71T, 2019ApJ...880...24T} linking the Pavo-Indus and Perseus-Pisces filaments.

\citet{2013AJ....146...69C} also identified a filament emanating from the Centaurus Cluster that splits into two branches. One passes through the Fornax Cluster and Eridanus cloud on its way to the Perseus-Pisces complex and is labelled For-PP at 5:00 of the Part-3 Perseus-Pisces video (animated Figure~\ref{fig:video_perseuspisces}).  The other passes through NGC~6868 and the Telescopium-Grus cloud \citep{1987nga..book.....T} on its way to the Perseus-Pisces region.  

The filaments discussed in this section can all be seen as fragments of an envelope bounding the Local Void \citep{2019ApJ...880...24T}.  The flow patterns out of the Local Void are nicely seen at 6:45 and 7:32 of the Part-3 Perseus-Pisces video (see animated Figure~\ref{fig:video_perseuspisces}). 

In the final scenes of the Part-1 Overview video (see animated Figure~\ref{fig:video_overview}), the envelope of Laniakea Supercluster is superimposed as defined by \citet{2014Natur.513...71T}.  Aside from the Arrowhead ``mini-supercluster" 
\citep{2015ApJ...812...17P} that is too inconsequential to be well proscribed by this orbit analysis, there are two neighboring structures that to a greater or lesser degree fall within the 8,000~\kms\ confine of this study: Perseus-Pisces and Great Wall.  It is seen that the putative boundary of Laniakea actually extends beyond our study region.  That outer area is mostly the domain of the Sculptor Void \citep{2019ApJ...880...24T}.  
At the interfaces between Laniakea and alternatively Perseus-Pisces and the Great Wall, the boundaries of separable gravitational wells lie in voids, albeit threaded with shearing filaments.

\section{Next Hubble time}
\label{sec:future}

\begin{figure*}[ht]
    \centering
    \includegraphics[width=0.9\textwidth]{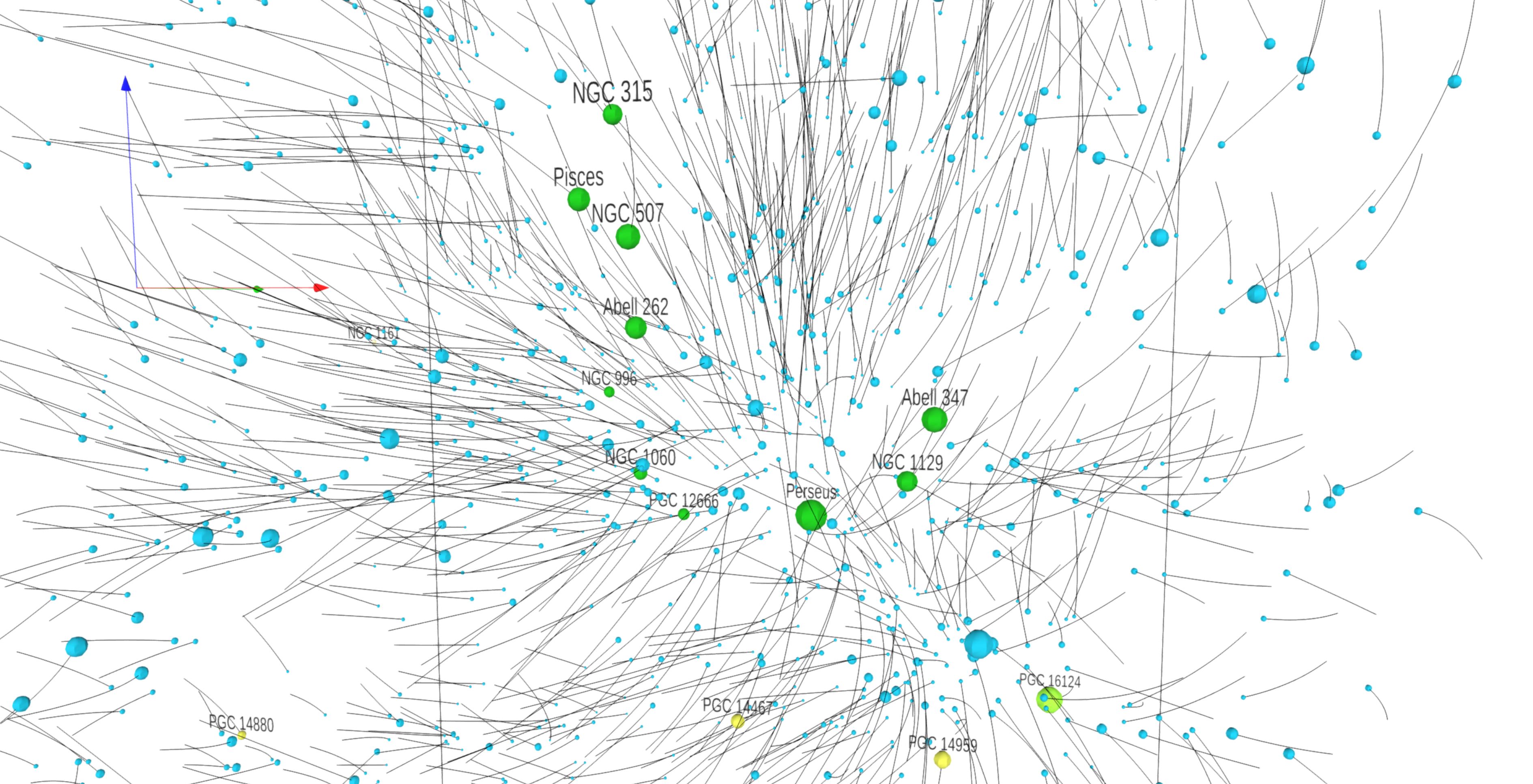}
    \caption{\textbf{Zoom visualization of the Perseus-Pisces region.} Start 4D
    interaction at \href{https://bit.ly/NAM8KPP}{https://bit.ly/NAM8KPP}. In this visualization of the n-body solution, the time evolution is run on the basis of one second for one Gyr.}
    \label{fig:Sketchfab_perseuspisces}
\end{figure*}

The orbits in our model can be computed into the future.  There is an asymmetry.  Information on mergers in the past is lost, with the consequences swallowed into the elements of our model constrained at the present.  By contrast, we can predict the occurrences of mergers in the future.  Any prediction is only as good as the representation of our orbits, so caution is warranted in specific cases.  Moreover, the elements are not the point-like billiard balls implicit to our model.  The elements are extended and composed of many particles, such that they will tend to coalesce through dynamical friction as their envelops overlap.  In our Part-4 Into the Future video (see animated Figure~\ref{fig:video_intothefuture}) and two interactive displays going into the future (see interactive Figures~\ref{fig:Sketchfab_greatattractor} and~\ref{fig:Sketchfab_perseuspisces}), elements pass through each other or dance in ways not to be believed.

The overarching picture for the future is boring.  Except in a small fraction of places, the velocities of attraction toward high density locations are overwhelmed by the cosmic expansion, made worse by $\Lambda$ as the expansion rate $H(t)$ never drops below $H_0 \Omega_{0,\Lambda}^{1/2}$. 
This basic truth is obscured by presentations in co-moving coordinates.  (See the Part-4 video in Figure~\ref{fig:video_intothefuture} at 0:53 to see the model evolve in physical coordinates.)  Most merging occurred in the distant past when entities were close to each other in physical space.  The onset of the influence of dark energy only accentuated the drift to boredom. 

Nevertheless, there will be some merging in the future.  We continued the computed paths in our $H_0 = 73$~\kms\ model until $a=2$, when the age of the universe will be 23.66~Gyr and $H = 64$.  It is interesting to consider a few cases where the events will be particularly substantial and the evidence is reasonably strong.  We will give attention to what is projected to happen along the Perseus-Pisces chain of clusters (see interactive Figure~\ref{fig:Sketchfab_perseuspisces}), the core of the Great Attractor (see interactive Figure~\ref{fig:Sketchfab_greatattractor}), and our immediate region from the Local Group to the Virgo Cluster.

Our numerical action orbit reconstruction emphatically confirms that the Perseus Cluster (Abell 426, Halo 200001 \citep{2015AJ....149..171T}) is the dominant attractor within the Perseus-Pisces filament, as seen in the interactive Figure~\ref{fig:Sketchfab_perseuspisces}.  The convergence of orbits toward this cluster, already apparent in orbits from the past, becomes more explicit with orbits into the future that pass through or loop around Perseus (see 1:50 and following in the Part-4 Into the Future video in the animated Figure~\ref{fig:video_intothefuture}).  There is a secondary gravitational basin along the filament involving the Pisces Cluster (NGC~410 Cluster: Halo 200005).  Evidently, there will be a collision between Pisces and the comparably large NGC~507 Cluster (Halo 200006) when the universe has an age of 22 Gyr (3:10 in the Part-4 Into-the-Future video Figure~\ref{fig:video_intothefuture}).  The rich Abell~262 (Halo 200003) and NGC~315 (Halo 200070) clusters are in the neighborhood. 

In the densest part of the Great Attractor region, there are two foci of convergence.  The nearer is around the Centaurus Cluster (Abell~3526).  In recognition of the two kinematic components to the traditional Centaurus Cluster, our modeling considers that there are two entities converging upon each other roughly today \citep{1986MNRAS.221..453L, 1997A&A...327..952S}.  The larger portion with dominant galaxy NGC~4696 (1PGC~ 43296, Halo 100003)\footnote{1PGC is the Principal Galaxies Catalog name \citep{1996A&A...311...12P} of the dominant galaxy in a group. Halo names are from \citet{2015AJ....149..171T}.} has observed velocities $\sim$3,100~\kms.  The smaller portion about dominant galaxy NGC~4709 (1PGC~43423, Halo 120003) has observed velocities $\sim$4,600~\kms.  The orbits of these two elements are merging by construction.  The pass through projected for the future will surely be damped by dynamical friction.  Meanwhile, other elements will have merged or will be on the verge of merging with Centaurus by twice the current age of the universe (Part-4 Into-the-Future video at 4:21 in the animated Figure~\ref{fig:video_intothefuture}).

The other focus of convergence (seen in the same video sequence) is the vicinity referred to by \citet{2013AJ....146...69C} as the `4 clusters strand'.  
Along the strand lie Abell~3565 (1PGC~48040, Halo 100027), Abell~3574 (1PGC~49025, Halo 100011), and Abell~S753 (1PGC~50100, Halo 100012).  Abell~3537 lies apart, but the cluster around NGC~5135 (1PGC~46974, Halo 100074) is a good fourth element.  
Over a doubling of the age of the universe, elements merge with these major clusters and the four draw closer in co-moving coordinates but they move apart in physical coordinates.  
A more refined model at the considerable distances of these clusters is needed to know their fate.  

Closer to home, the Virgo Cluster is easier to study \citep{2017ApJ...850..207S}.  Many distances are known with high accuracy and the associated galaxies consequently have well determined peculiar velocities.  Tip of the red giant branch contributions with 5\% accuracies to about 500 galaxies enrich the analysis.  Tidal influences are well described on relevant scales. 

Infall and accretion into Virgo are seen following 5:45 in the Part-4 Into-the-Future video (Figure~\ref{fig:video_intothefuture})\footnote{The slingshot ejection of 3 elements including M61 and NGC~4636 at $\sim$17~Gyr are artifacts of the point-like nature of the model constituents.  It is expected that these elements will be consumed by the cluster.}.  We see a great deal of activity because the coverage is dense; the situation must be similar about other rich clusters at greater distance.  The focus of the current study is on large scales, so we defer to the more local numerical action model by \citet{2017ApJ...850..207S} for Virgo infall details. To summarize from that study, the mass within the collapsed cluster was found to be $6.3\pm0.8\times10^{14}~\Msun$ and the surface separating infall from expansion at a=1 (encompassing the galaxies that reach the cluster by $a=2$) is roughly triaxial, with radial dimensions 6.6~Mpc along the axis of current major infall (through the Virgo Southern Extension), 7.9~Mpc toward the relative voids at the Supergalactic poles, and an intermediate 7.2 Mpc on the axis aligned with our sight-line.  The mass in groups within turnaround including the Virgo Cluster is $8.3\times10^{14}~\Msun$.

The video pauses at a=1 to illustrate the view from our position today.  The concentration of pink elements that represent major members of the Virgo Southern Extension are found to be a transient occurrence today of elements from diverse origins in our model. 

The video leads us back home at 7:08.  We live in a boring place, as is common in the universe today.  The merger between the Milky Way and Andromeda anticipated in 6~Gyr \citep{2012ApJ...753....8V} is concealed here by the union of the two in the Local Group entity in our model.  The most exciting local event predicted is the merger between M83 and Centaurus~A, combining into the most massive halo within 8~Mpc.  We are trending toward this collapsing region in co-moving coordinates but will never be captured.

\section{Summary}
\label{sec:summary}

Most peculiar velocity studies invoke linear perturbation theory.  
{\it Cosmicflows-3} distance and velocity information \citep{2016AJ....152...50T} and 2MRS, the 2MASS redshift survey to $K_s=11.75$ \citep{2012ApJS..199...26H}, have been used in linear regime studies \citep{2019MNRAS.488.5438G, 2021arXiv210207291L, 2021MNRAS.505.3380H} and there have been studies with variant samples \citep{2015MNRAS.450..317C, 2019A&A...625A..64J, 2021MNRAS.502.3456K}.

However in regions of higher density, like the Great Attractor core of Laniakea Supercluster or the Perseus-Pisces complex, and in large voids, the linear approximation is inappropriate.  Orbit reconstructions from the Numerical Action Method \citep{1989ApJ...344L..53P, 2001ApJ...554..104P, 1995ApJ...454...15S, 2017ApJ...850..207S} provide a granular description of peculiar velocities up to the point of shell crossing.  Upon collapse, orbits become too confused to be recovered, so the entities whose orbits we follow are the virial / second turnaround domain of halos (be these called `groups' or `individual galaxies').  

The 9,719 entities drawn from the 2MRS $K_s=11.75$ catalog provides a good description of the distribution of old stars, expected to correlate with mass, over the unobscured 91\% of the volume within the 8,000~\kms\ limit of this study.  The grouping into turnaround domains was discussed by \citet{2015AJ....149...54T, 2015AJ....149..171T}, where justification is given for the linkage between observed light and underlying mass described by Eq.~\ref{eq:mass-to-light} and found adequate in the more local numerical action study by \citet{2017ApJ...850..207S}.  Somewhat lower halo $M/L_K$ values are preferred from the present analysis, as illustrated graphically as a function of $H_0$ in Fig.~\ref{fig:alongmin}.

Mass in the 9\% of the sky behind the Milky Way is poorly established.  We assign mass to known entities sufficient to maintain average density in the zone of obscuration but the orbits of these and neighboring objects should not be trusted.  Likewise, our description of external tidal influences with 44 major clusters at $8,000-16,000$~\kms\ is crude.  Were we to repeat this exercise, we might use the description of matter in this velocity regime given by \citet{2021arXiv210207291L}.

The other essential ingredients for this analysis are the distance measures and inferred peculiar velocities provided by {\it Cosmicflows-3} \citep{2016AJ....152...50T}.  Although individual errors are often large, averaging within groups provides a corpus with small statistical uncertainties. 

Orbits of our $\sim$10,000 constituents are constrained by the numerical action methodology \citep{1989ApJ...344L..53P} to give end-point velocities and distances consistent with observations.  However,  we know both from observations of the Ly$\alpha$ forest and from numerical simulations that there must be significant mass outside the collapsed entities whose orbits we follow.  Our simplistic method to account for this component is to provide a uniform density described by the parameter $\Omega_\mathrm{IHM}$.
The total local density is the sum of the contributions in halos and in the interhalo medium: $\Omega_\mathrm{local} = \Omega_\mathrm{halos} + \Omega_\mathrm{IHM}$.

We find optimal models, assuming the Planck Satellite acoustic angular scale constraint: $\Omega_m h^3 = \xi_P = 0.096345$ \citep{2020A&A...641A...6P}.  We find a good fiducial model with 
$\Omega_\mathrm{local} \simeq \Omega_m$, with $H_0 \simeq 73.0$~\kmsMpc\ and $\Omega_m \simeq 0.248$, with half the mass in the observed bound mass of galaxies, groups, and clusters.  The other half would reside in a non-luminous medium lurking beyond these entities.

Good models with other values of $H_0$ can be found if the constraint of $\delta=0$ is relaxed. 
We find, both by numerical models of the 8,000 \kms\ region with alternative values of $\Omega_\mathrm{IHM}$, and by analytical models, how $\delta$ varies with global $H_0$ along the best fit valley in \meanchi. 
Models with acceptable $\chi$ values can be found with $H_0$ almost as low as the range of Planck and other CMB determinations, but these cases require that we live in an unusual underdensity ($\sim$30\% to reach as low as $H_0=69$) over a volume of at least 100 Mpc in radius.  

From observations, on the scale of 30 Mpc, we live in the overdense region of the Local Supercluster and its appendage to the Great Attractor complex.  
 On larger scales of $\sim$100 Mpc, some $K$-band and x-ray based studies suggest \citep{2013ApJ...775...62K, 2020A&A...633A..19B} that we may live in a substantial hole. These studies are difficult to interpret, though, because the concentration of stars and, particularly, thermal x-ray emission arising from massive clusters, are likely biased tracers of mass \citep{1984ApJ...284L...9K}.

The underdensities of the most immediate Local and Hercules voids \citep{2019ApJ...880...24T} are reasonably offset by the bounding Perseus-Pisces and Great Wall structures.  However, a very large fraction of the volume south of the Milky Way is the domain of the Sculptor Void \citep{1991MNRAS.248..313K, 2019ApJ...880...24T}, leaving room for doubt as to whether our study region is underdense.  
Accordingly, there could be a preference for a lower value of $H_0$ that accommodates an outflow. 
Considerations derived from simulations, discussed in \S\ref{sec:simulations}, favor the value of $H_0 \sim$72 \kms\ because of the expectation of the fraction of mass outside of the massive halos of about 41\%.  However there must be reservations in a comparison of our model of discrete halos on a smooth IHM pedestal with the complexity of the real universe.
Other external information regarding the Hubble Constant obtained from the far field from Type~Ia supernovae, consistent with our zero-point calibration \citep{2014ApJ...792..129N,2021ApJ...908L...6R}, place limits on $H_0$ in the $\sim$73 range.  
Using a sample of 1300 SNIa, \citet{2019ApJ...875..145K} find no evidence of a change in Hubble Constant values across the redshift range $0.02<z<0.15$.

Turning to the orbits found in our best models, the most evident features are the flow patterns that formed the three main concentrations of galaxies: the core of Laniakea, the Perseus-Pisces spine, and the nearest sector of the Great Wall.  Shears in the flows occur in the voids: in the Local Void between Laniakea and Perseus-Pisces, and in the Hercules Void between Laniakea and the Great Wall.  The largest of the nearby voids, the Sculptor Void, extends well beyond our study region, creating a division between Laniakea and the South Pole Wall \citep{2020ApJ...897..133P}.

Drilling down in detail, we document the dominant flow downward at all positive $SGZ$ over the Laniakea region, including the substantial coherent flow along the Pavo-Indus component of Laniakea.  The focus of the flow within the Laniakea core, the Great Attractor \citep{1987ApJ...313L..37D}, is not well specified.  At best, we place it in the proximity of what we call the 4-clusters region \citep{2013AJ....146...69C}.  The Centaurus and Norma clusters are important constituents somewhat apart.

The flow histories are best represented by the accompanying videos and Sketchfab interactive displays.  It is interesting to take note of the filamentary connections between the densest regions, particularly those between Laniakea and Perseus-Pisces, such as the Centaurus-Puppis-PP filament.  

Of course, while sectors of space can be parsed in co-moving coordinates \citep{2014Natur.513...71T}, the expansion of space in physical coordinates is overwhelmingly dominant on all but very local scales in a flat $\Lambda$CDM universe.  
The large scale structures continue to expand in one or two dimensions becoming ever thinner.  
For the most part, halos have stopped growing.  However, our orbits, to the degree that they can be individually believed, as extended in time until the scale of the universe doubles, and in interesting regions, such as around the Perseus, Centaurus, and Virgo clusters, present scenarios of a few merger events that will occur in the far future.

\section{acknowledgements}

Above all, we recognise the critical contributions of P.J.E. (Jim) Peebles to the development of the numerical action methodology.  Jim modestly demurred from being a co-author of this paper although it could be viewed as a culmination of his inspiration.  Any deficiencies in the exploitation rest with the authors.  We also thank our colleague Yehuda Hoffman who provided a file of the 3D density field from an as yet unpublished quasi-linear modeling of {\it Cosmicflows-3} distances.  This material is seen as the iso-density features of the videos and interactive displays.  Funding for the {\it Cosmicflows} project has been provided by the US National Science Foundation grant AST09-08846, the National Aeronautics and Space Administration grant NNX12AE70G, and multiple awards to support observations with Hubble Space Telescope through the Space Telescope Science Institute.  The resources of the NASA-IPAC Extragalactic Database and HyperLEDA hosted in Lyon, France have been indispensable. 

\bibliography{NAM8k}
\bibliographystyle{aasjournal}

\end{document}